\documentclass[pre,onecolumn,showpacs,superscriptaddress,aps]{revtex4}

\usepackage{amsfonts,xfrac}
\usepackage{amssymb}
\usepackage{physymb}
\usepackage{graphicx}
\usepackage{dcolumn}
\usepackage{dsfont}
\usepackage{times}
\usepackage{epsfig}
\usepackage{multirow} 
\usepackage{hhline} 
\usepackage{makecell}
\usepackage{diagbox}
\usepackage{booktabs}
\usepackage[hang,nooneline]{subfigure}                         
\newcounter{fig}   

\newcommand{\R}{{\mathbb R}}

\newcommand{\Z}{{\mathbb Z}}

\DeclareSymbolFont{matha}{OML}{txmi}{m}{it}
\DeclareMathSymbol{\varv}{\mathord}{matha}{118}

\begin{document}

\title{Time-Periodic Solutions of Driven-Damped Trimer Granular Crystals}

\author{E. G. Charalampidis\footnote{Email: charalamp@math.umass.edu}}
\affiliation{School of Civil Engineering, Faculty of Engineering, Aristotle %
University of Thessaloniki, Thessaloniki 54124, Greece\\}
\affiliation{Department of Mathematics and Statistics, University of Massachusetts, Amherst, MA 01003-4515, USA}
\author{F. Li \footnote{Email: lif@ciomp.ac.cn}}
\affiliation{Aeronautics \& Astronautics, University of Washington, Seattle, WA 98195-2400, USA}
\author{C. Chong\footnote{Email: cchong@ethz.ch}}
\affiliation{Department of Mathematics and Statistics, University of Massachusetts, Amherst, MA 01003-4515, USA}
\affiliation{Department of Mechanical and Process Engineering (D-MAVT), \\ %
Swiss Federal Institute of Technology (ETH), 8092 Z\"urich, Switzerland}
\author{J. Yang \footnote{Email: jkyang@aa.washington.edu}}
\affiliation{Aeronautics \& Astronautics, University of Washington, Seattle, WA 98195-2400, USA}
\author{P. G. Kevrekidis\footnote{Email: kevrekid@math.umass.edu}}
\affiliation{Department of Mathematics and Statistics, University of Massachusetts, Amherst, MA 01003-4515, USA}

\affiliation{Center for Nonlinear Studies and Theoretical Division, Los Alamos
National Laboratory, Los Alamos, NM 87544, USA}

\date{\today}

\begin{abstract}
In this work, we consider time-periodic structures of trimer granular 
crystals consisting of alternate chrome steel and tungsten carbide spherical
particles yielding a spatial periodicity of three. The configuration at 
the left boundary is driven by a harmonic in-time actuation with given 
amplitude and frequency while the right one is a fixed wall. 
Similar to the case of a dimer chain, the combination of dissipation, 
driving of the boundary, and intrinsic nonlinearity leads to complex 
dynamics. For fixed driving frequencies in each of the spectral gaps,  
we find that the nonlinear surface modes and the states dictated
by the linear drive collide in a saddle-node bifurcation as the driving
amplitude is increased, beyond which the dynamics of the system become
chaotic. While the bifurcation structure is similar for solutions within
the first and second gap, those in the first gap appear to be less robust.
%
We also conduct a continuation in driving frequency, where it is apparent
that the nonlinearity of the system 
results in a complex bifurcation diagram, involving 
an intricate set of loops of branches, especially 
within the spectral gap. 
%
The theoretical findings are qualitatively corroborated 
by the experimental full-field
visualization of the time-periodic structures. 
\end{abstract}


\maketitle

\section{Introduction} \label{sec:intro}
Granular chains, which consist of closely packed arrays of particles that
interact elastically, have proven over the last several decades to be an 
ideal testbed to theoretically and experimentally study novel principles 
of nonlinear dynamics \cite{Nesterneko_book,sen08,Theocharis_rev}. Examples
include, but are not limited to, solitary waves \cite{Nesterneko_book,sen08,coste,pik} and 
dispersive shocks \cite{herbold,molin}, as well as
bright and dark discrete breathers \cite{Theo2009,Theo2010,Nature11,hooge12,hooge13,dark,dark2}. 
Beyond such fundamental aspects, their extreme tunability makes granular 
crystals relevant for numerous applications such as shock and energy absorbing 
layers~\cite{dar06,hong05,fernando,doney06}, actuating devices \cite{dev08}, 
acoustic lenses \cite{Spadoni}, acoustic diodes \cite{Nature11} and switches \cite{Li_switch}, as well as sound 
scramblers \cite{dar05,dar05b}.

Our emphasis in the present work will be on coherent nonlinear waveforms
that are time-periodic. A special instance of this is when the spatial 
profile is localized, in which case the structure is termed a discrete 
breather. The study of discrete breathers has been a topic of intense
theoretical and experimental interest during the 25 years 
since their  theoretical inception, as has been 
summarized, e.g., in \cite{Flach2007}. 
The broad and diverse span of fields where such structures have 
been of interest includes, among others, 
optical waveguide arrays or photorefractive 
crystals~\cite{moti}, micromechanical cantilever arrays~\cite{sievers},
Josephson-junction ladders~\cite{alex},
layered antiferromagnetic crystals~\cite{lars3}, halide-bridged transition
metal complexes~\cite{swanson}, dynamical models of the DNA double 
strand \cite{Peybi} and Bose-Einstein condensates 
in optical lattices~\cite{Morsch}.

In Fermi-Pasta-Ulam type settings (which are intimately connected
to the realm of precompressed granular crystals),
it was proven in \cite{James01,James03}  (see also the discussion in
\cite{Flach2007}) that small amplitude discrete breathers are absent
in spatially homogeneous (i.e. monoatomic) chains. 
Instead, dark such states (those on top of a non-vanishing 
background) have been found therein~\cite{dark,dark2}.
It is for that reason that 
the first theoretical and experimental investigations of breathers
with a vanishing background (i.e., bright breathers) have taken place
in granular chains with some degree
of spatial heterogeneity, which plays a critical role in the emergence
of such patterns.
Examples include chains with defects \cite{Theo2009,Nature11}
(see also for recent experiments~\cite{man})
and a spatial periodicity of two (i.e. dimer lattices) \cite{Theo2010,hooge12}. 
Further recent experimental works 
explored solitary
waves in trimers and higher periodicity chains~\cite{mason1,mason2},
as well as the linear dispersion properties of
such chains~\cite{Tun_Band_gaps}.
Motivated by these works,
a theoretical study of breathers in granular crystals with higher order 
spatial periodicity (such as trimers and quadrimers) was recently conducted
in \cite{hooge13}. Therein, it was demonstrated that breathers with a frequency
in the highest gap appear to be more robust than their counterparts with 
frequency in the lower gaps. 

The goal of the present work is the systematic study of time-periodic 
solutions (including breathers) of trimer granular crystals with frequency
in the first or second gap, as well as 
in the acoustic and optical bands. In particular,
we investigate the robustness of the breathers experimentally using a full-field
visualization technique based on laser Doppler vibrometry. This is a 
significant improvement over the aforementioned experimental
observation of bright breathers \cite{Theo2010,Nature11,hooge12} where force sensors
are placed at isolated particles within the granular chain, which does not
allow a full-field representation of the breather. We complement this study
with a detailed theoretical probing of the more realistic damped-driven variant
of the pertinent model. Our extensive analysis of such modes consists
of the study of their family under 
continuations in both the amplitude and the frequency
parameters of the external drive and a detailed comparison of the
findings between numerically exact (up to a prescribed tolerance)
periodic orbit solutions and experimentally traced counterparts thereof.

The paper is structured as follows. In Secs.~\ref{exp} and \ref{thr}, we 
describe the experimental and theoretical set-ups respectively. The main
results are presented in Sec.~\ref{main}, where time-periodic solutions 
with frequency in the first/second gaps and in the spectral bands are
studied in both of these setups and compared accordingly. 
Finally, Sec.~\ref{theend} provides our conclusions and 
discusses a number of future challenges.

\section{Experimental Setup} \label{exp}

\begin{figure}[!pt]
\begin{center}
\vspace{-0.2cm}
\mbox{\hspace{-0.2cm}
\includegraphics[height=.25\textheight, angle =0]{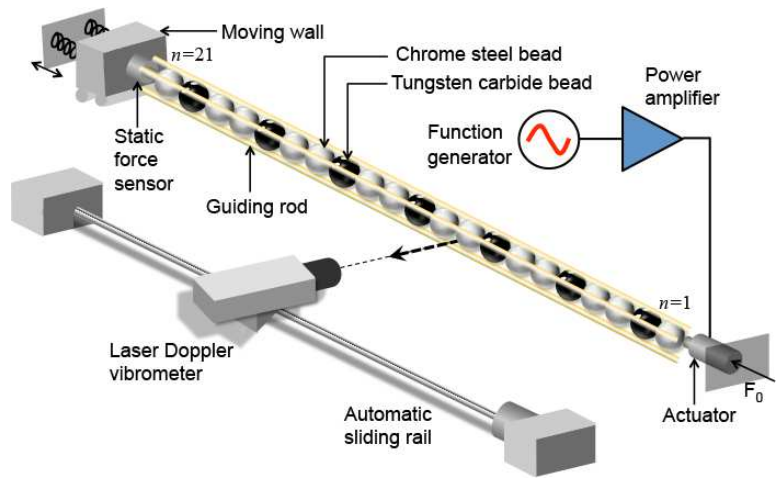}
}
\end{center}
\caption{(Color online) Experimental setup of a 21-particle granular chain composed of chrome steel and tungsten carbide beads in a 2:1 ratio. Actuation and sensing systems based on a piezoelectric force sensor and a laser Doppler vibrometer are also illustrated.
}
\label{expSetup}
\end{figure}

Figure 1 shows a schematic of the experimental setup consisting of a granular chain and a laser Doppler vibrometer. In this study, we consider a granular chain composed of $N=21$ spherical beads. The beads are made out of chrome steel (S: gray particles in Fig. \ref{expSetup}) and tungsten 
carbide (W: black particles) materials. See Table~\ref{gc_params} for nominal values of the material 
parameters used hereafter. The granular chain has a spatial periodicity of three particles, and each unit cell 
follows the pattern of a 2:1 trimer: $\textrm{S}-\textrm{W}-\textrm{S}$. The spheres are supported by four polytetrafluoroethylene (PTFE) rods, which allow axial vibrations of particles with minimal friction, while restricting their lateral movements. The granular chain is compressed by a moving wall at one end of the chain that applies static force in a controllable manner via a linear stage (see Fig. \ref{expSetup}). We measure the pre-applied static force ($F_0$ = 10 N in this study) by using a static force sensor mounted on the moving wall. We assume that this moving wall is stationary throughout our analysis, since it exhibits orders-of-magnitude larger inertia compared to the particle's mass.

The granular chain is driven by a piezoelectric actuator positioned on the other side of the chain. We impose actuation signals of chosen amplitude and frequency through an external function generator and a power amplifier. The dynamics of individual particles are scoped by a laser Doppler vibrometer (LDV, Polytec, OFV-534), which is capable of measuring particles' velocities with a 
resolution of 0.02 $\mu$m/s/Hz$^{1/2}$. The LDV scans the granular chain through the automatic sliding rail and measures the vibrational motions of 
each particle three times for statistical purposes. We obtain the full-field map of the granular chain's dynamics by synchronizing and reconstructing the acquired data.

\begin{table}[pht]
\caption{Parameter values of the trimer granular chain $(\textrm{S}-\textrm{W}-\textrm{S})$.}
\centering
\begin{tabular}{ | c | c | c | c | c | }
\hline
\theadfont\diagbox[width=14em]{Material}{Parameter}   &  $E \,[\textrm{N}/\textrm{m}^{2}]$   & $r \,[\textrm{m}]$ & $\nu$ & $ \rho\,[\textrm{kg}/\textrm{m}^{3}]$ \\  \hline \hline
   \textrm{Chrome steel (S)}      &  $200\times10^{9}$  &  $9.525\times10^{-3}$  &   $0.3$   &    $7780$      \\ \hline
   \textrm{Tungsten Carbide (W)}  &  $628\times10^{9}$  &  $9.525\times10^{-3}$  &   $0.28$  &    $14980.64$  \\      
\hline
\end{tabular}
\label{gc_params}
\end{table}

\section{Theoretical Setup} \label{thr}
The equation we use to model the experimental setup 
is a Fermi-Pasta-Ulam-type lattice with a Hertzian potential \cite{Nesterneko_book} leading to:
\begin{equation}
\ddot{u}_{n} = \frac{A_{n-1}}{M_{n}}\left[\delta_{0,n-1} + u_{n-1} - u_{n}\right]^{3/2}_{+} %
              - \frac{A_{n}}{M_{n}}\left[\delta_{0,n} + u_{n} - u_{n+1}\right]^{3/2}_{+} %
              - \frac{\dot{u}_{n}}{\tau}, 
\label{gc_start}              
\end{equation}
where $n=1,\dots,N$, 
$u_n=u_n(t)\in\R$ is the displacement of the $n$-th bead from 
its equilibrium position at time $t\in\R$, $M_n$ is the mass of the $n$-{th} 
bead, $\delta_{0,n}$ is a precompression factor induced by the static
force $F_{0}=A_{n} \delta_{0,n}^{3/2}$ and the bracket is 
defined by $[x]_{+}=\max(0,x)$. The $3/2$ power accounting for
the nonlinearity of the model is a result of the
sphere-to-sphere contact, i.e., the so-called Hertzian contact \cite{Hertz}. The form of the dissipation is a dash-pot, 
which has been utilized in the context of granular crystals in several previous works \cite{Nature11,dark2}.
The strength of the dissipation is captured by the parameter $\tau$, which serves as the sole parameter used
to fit experimental data ($\tau = 2.1$ ms in this study).
The elastic coefficient $A_n$ depends on the interaction of bead $n$ with bead $n+1$ and for spherical
point contacts has the form \cite{Johnson_contact} 
\begin{equation}
A_{n} = \frac{4 E_{n}E_{n+1}\sqrt{\frac{r_{n}r_{n+1}}{\left(r_{n}+r_{n+1}\right)}}}%
{3E_{n+1}\left(1-\nu_{n}^2\right) + 3 E_{n}\left(1-\nu_{n+1}^2\right)  },
\end{equation}
where $E_{n}$, $\nu_{n}$ and $r_n$ are the Young's modulus, Poisson's
ratio and the radius, respectively, of the $n$-th bead. The left 
boundary is an actuator and the right one is kept fixed, i.e.,
\begin{equation}
u_{0}(t) = a \cos(2\pi f_b t), \qquad u_{N+1}(t) = 0,
\label{actuator}
\end{equation}
where $a$  and $f_{b}$ represent the driving amplitude and frequency, 
respectively. Note that at the boundaries (i.e., $n=0$ and $n=N+1$),
we consider flat surface-sphere contacts while the same material properties
are assumed; thus, e.g., the elastic coefficient at the left boundary takes the form 
\begin{equation}
A_{0}=\frac{2E_{1}\sqrt{r_{1}}}{3\left(1-\nu_{1}^{2}\right)}.
\end{equation}

\subsection{Linear regime and dispersion relation}
\begin{figure}[!pt]
\begin{center}
\vspace{-0.2cm}
\mbox{\hspace{-0.2cm}
\subfigure[][]{\hspace{-0.2cm}
\includegraphics[height=.18\textheight, angle =0]{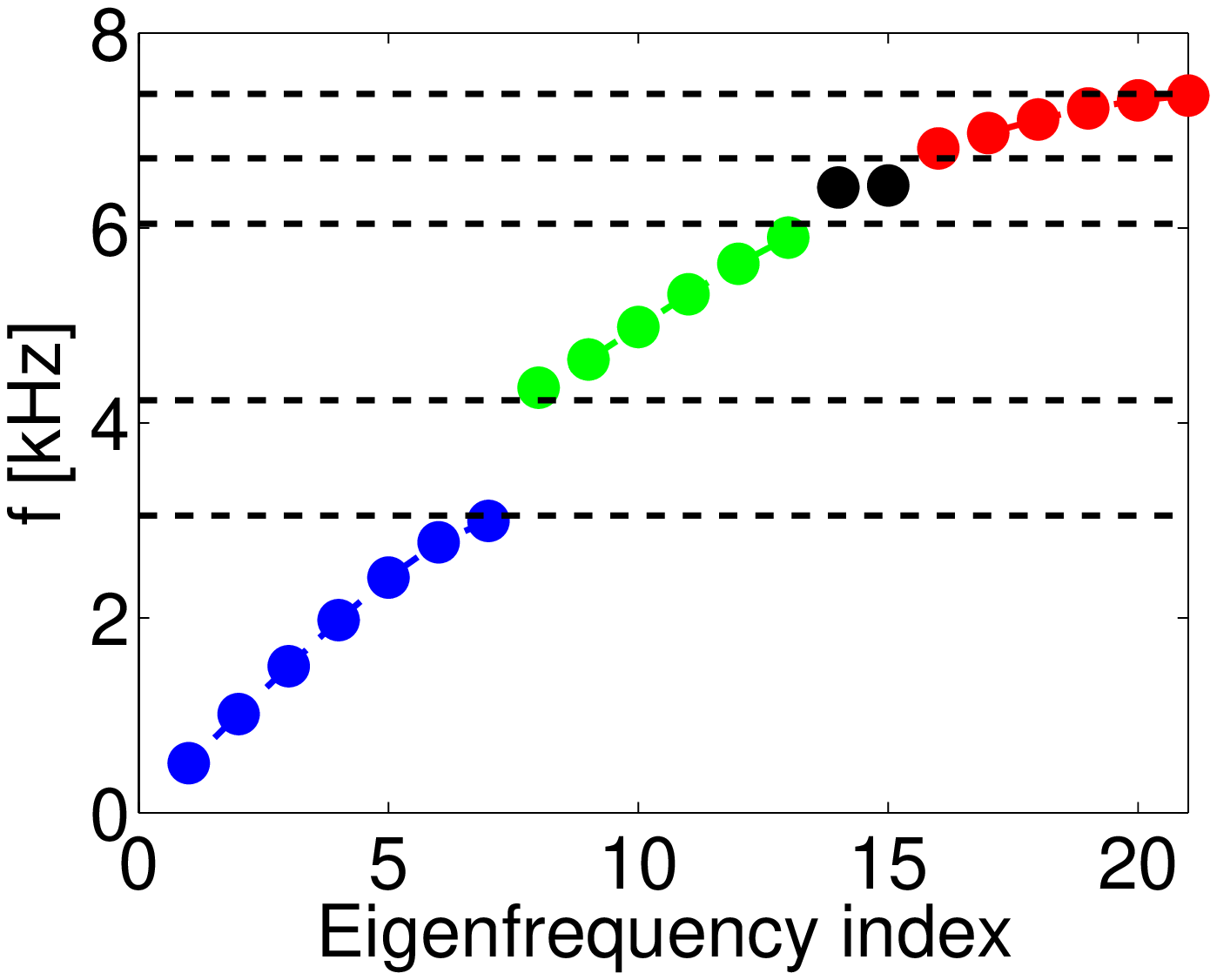}
\label{fig1a}
}
\subfigure[][]{\hspace{-0.2cm}
\includegraphics[height=.18\textheight, angle =0]{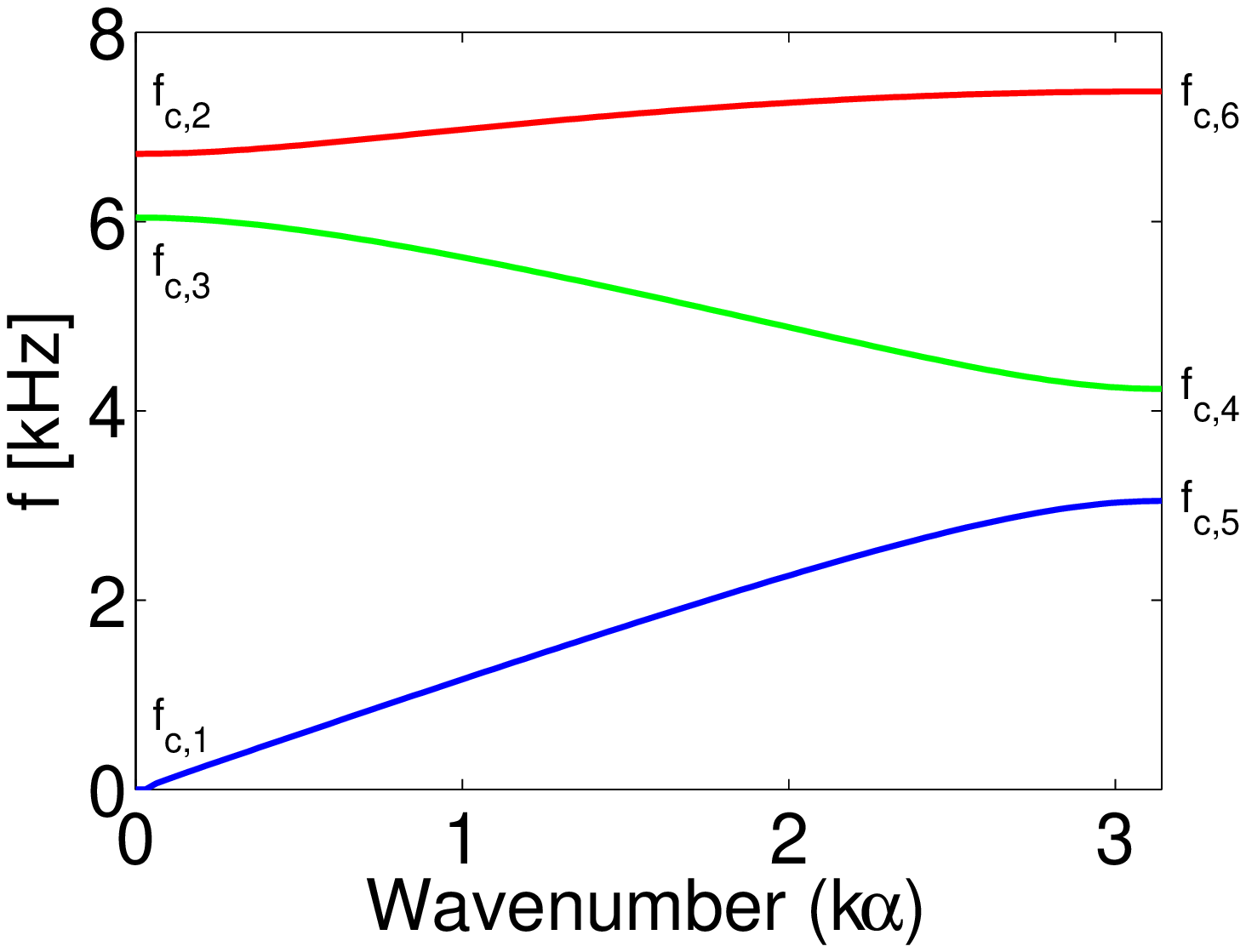}
\label{fig1b}
}
}
\end{center}
\caption{(Color online) Eigenfrequency spectrum of the 
trimer granular chain for parameter values given in Table~\ref{gc_params} 
and $\tau = 2.1$ ms. \textbf{(a)}  Spectrum corresponding to  Eq.~(\ref{eigv_prob_linear}) 
for $N=21$ beads. Note that the horizontal dashed black lines correspond to
the cut-off frequencies listed in Table~\ref{cut_off_freq_values}. \textbf{(b)}
Numerically obtained frequencies of the infinite lattice using Eq.%
~(\ref{dispersion_relation}) as a function of the wavenumber $(k\alpha)$. 
}
\label{fig1}
\end{figure}

Assuming dynamic strains that are small relative to the static 
precompression, i.e.,
%
\begin{equation}
\frac{|u_{n}-u_{n+1}|}{\delta_{0,n}}\ll 1
\label{lin_limit}
\end{equation}
Eq.~(\ref{gc_start}) can be well approximated by its linearized form
\begin{equation}
\ddot{u}_{n} = \frac{K_{n-1}}{M_{n}}\left(u_{n-1} - u_{n}\right) - \frac{K_{n}}{M_{n}}\left(u_{n} - u_{n+1}\right)- \frac{\dot{u}_{n}}{\tau},
\label{gc_linear}
\end{equation}
with 
$ {K_{n}=3/2 \, A_{n}\delta_{0,n}^{1/2}}$ 
corresponding 
to the linearized stiffnesses. Subsequently, Eq.~(\ref{gc_linear}) can be 
converted into a system of $2N$ first-order equations and written conveniently
in matrix form as 
\begin{equation}
\dot{\mathbf{Y}}=\mathcal{A}\,\mathbf{Y}, 
\label{gc_linear_matrix}
\end{equation}
with $\mathbf{Y}=\Big(u_{1},\dots,u_{N},\varv_{1}\left(=\dot{u}_{1}\right),\dots,%
                                  \varv_{N}\left(=\dot{u}_{N}\right)\Big)^{T}$ and
\begin{equation}
\mathcal{A} =
\begin{pmatrix}
 \mathbf{O} & \mathbf{I} \\
\mathcal{C} & (-1/\tau)\,\mathbf{I}
\end{pmatrix},
\label{matrix_linear_A}
\end{equation}
where $\mathbf{O}$ and $\mathbf{I}$ represent the $N\times N$ zero and
identity matrices, respectively, and the $N\times N$ (tridiagonal)
matrix $\mathcal{C}$ is given by
\begin{equation}
 \mathcal{C} =
\begin{pmatrix}
-\frac{K_{0}+K_{1}}{M_{1}} & \frac{K_{1}}{M_{1}}        &         0               &            \cdots                &               0             \\
  \frac{K_{1}}{M_{2}}      & -\frac{K_{1}+K_{2}}{M_{2}} & \frac{K_{2}}{M_{2}}     &                                  &            \vdots           \\
             0             &           \ddots           &      \ddots             &            \ddots                &               0             \\
          \vdots           &                            & \frac{K_{N-2}}{M_{N-1}} & -\frac{K_{N-2}+K_{N-1}}{M_{N-1}} & \frac{K_{N-1}}{M_{N-1}}      \\
             0             &           \cdots           &         0               &   \frac{K_{N-1}}{M_{N}}          & -\frac{K_{N-1}+K_{N}}{M_{N}}
\end{pmatrix}.
\label{matrix_C}
\end{equation}
Note that in the above linearization 
we 
applied fixed boundary conditions at both ends of the chain, i.e.,
$u_{0}=u_{N+1}=0$ (we account for the actuator in the following subsection). 
This equation is solved by $Y_{n}=y_{n}e^{i\omega t}$, where $\omega$ corresponds
to the angular frequency (with $\omega=2\pi f$) and where,
%
\begin{equation}
\mathcal{A}\,\mathbf{y} = \lambda \mathbf{y}, 
\label{eigv_prob_linear}
\end{equation}
with $\left(\lambda,\mathbf{y}\right)$ corresponding to the eigenvalue-eigenvector
pair, while $\lambda=i\omega$ and $\mathbf{y}=\left(y_{1},\dots,y_{2N}\right)^{T}$.
The eigenfrequency spectrum $\left(f=-i \frac{\lambda}{2\pi}\right)$ using the values
of Table~\ref{gc_params} and $\tau = 2.1\,\textrm{ms}$ is shown in Fig.~\ref{fig1a}.

On the other hand, the dispersion relation for the trimer chain 
configuration within an infinite lattice can be obtained using 
a Bloch wave ansatz \cite{Tun_Band_gaps}. We first re-write Eq.~(\ref{gc_linear})
in the following convenient form
%
%
\begin{subequations}
 \begin{eqnarray}
\ddot{u}_{3j-2}&=&\frac{K_{3}}{M_1}\left(u_{3j-3}-u_{3j-2}\right)-\frac{K_{1}}{M_{1}}\left(u_{3j-2}-u_{3j-1}\right)-\frac{\dot{u}_{3j-2}}{\tau},  \\
\ddot{u}_{3j-1}&=&\frac{K_{1}}{M_{2}}\left(u_{3j-2}-u_{3j-1}\right)-\frac{K_{1}}{M_{2}}\left(u_{3j-1}-u_{3j}\right)-\frac{\dot{u}_{3j-1}}{\tau},  \\
\ddot{u}_{3j}  &=&\frac{K_{1}}{M_{1}}\left(u_{3j-1}-u_{3j}\right)-\frac{K_{3}}{M_{1}}\left(u_{3j}-u_{3j+1}\right)-\frac{\dot{u}_{3j}}{\tau},
\end{eqnarray}
\label{gc_linear_split}
\end{subequations}
with $j\in\Z^{+}$, where we made use of the spatial periodicity of the trimer lattice, i.e., 
$M_{3j}=M_{3j-2}\equiv M_{1}(=M_{3})$ and $K_{3j-2}=K_{3j-1}\equiv K_{1}(=K_{2})$,
and $M_{3j-1}\equiv M_{2}$ and $K_{3j}\equiv K_{3}$.  
The Bloch wave ansatz has the form,
%
\begin{subequations}
\begin{eqnarray}
u_{3j-2} = u_{0}e^{i\left(k\alpha j + \omega t\right)}, \\
u_{3j-1} = v_{0}e^{i\left(k\alpha j + \omega t\right)}, \\
  u_{3j} = w_{0}e^{i\left(k\alpha j + \omega t\right)},
\end{eqnarray}
\label{trav_wave_ansatze}
\end{subequations}
where $u_{0}, v_{0}$ and $w_{0}$ are the wave amplitudes, $k$ is the wavenumber
and $\alpha$ is the size (or the equilibrium length) of one unit cell of the 
lattice. Substitution of Eqs.~(\ref{trav_wave_ansatze}) into Eqs.~(\ref{gc_linear_split})
yields
\begin{equation}
\begin{pmatrix}
\omega^{2} - \frac{K_{1}+K_{3}}{M_{1}}-i\frac{\omega}{\tau} &                \frac{K_{1}}{M_{1}}                   & \frac{K_{3}}{M_{1}} e^{-i k \alpha}                      \\
                \frac{K_{1}}{M_{2}}                         & \omega^{2}-\frac{2K_{1}}{M_{2}}-i\frac{\omega}{\tau} & \frac{K_{1}}{M_{2}}                                       \\
           \frac{K_{3}}{M_{1}}e^{ik\alpha}                  &                 \frac{K_{1}}{M_{1}}                  & \omega^{2}-\frac{K_{1}+K_{3}}{M_{1}}-i\frac{\omega}{\tau}
\end{pmatrix}
\begin{pmatrix}
u_{0} \\ v_{0} \\ w_{0}
\end{pmatrix}
=\begin{pmatrix}
  0 \\0 \\ 0
 \end{pmatrix}.
 \label{dispersion_matrix_system}
\end{equation}
The non-zero solution condition of the matrix-system (\ref{dispersion_matrix_system}),
or dispersion relation (i.e., the vanishing of the determinant of the
above homogeneous linear system), has the form
\begin{eqnarray}
&&M_{1}^{2}M_{2}\omega^{6}-i\frac{3M_{1}^{2}M_{2}}{\tau}\omega^{5}%
-M_{1}\left[2\left[K_{3}M_{2}+K_{1}\left(M_{1}+M_{2}\right)\right]+\frac{3M_{1}M_{2}}{\tau^{2}}\right]\omega^{4}\nonumber \\
&&+i\frac{M_{1}}{\tau}\left[4\left[K_{3}M_{2}+K_{1}\left(M_{1}+M_{2}\right)\right]+\frac{M_{1}M_{2}}{\tau^{2}}\right]\omega^{3}%
-4K_{1}^{2}K_{3}\sin^{2}{\left(\frac{k\alpha}{2}\right)} \nonumber \\
&&+\left[K_{1}\left(K_{1}+2K_{3}\right)\left(2M_{1}+M_{2}\right)+\frac{2M_{1}}{\tau^{2}}\left[K_{3}M_{2}+K_{1}\left(M_{1}+M_{2}\right)\right]\right]\omega^{2} \nonumber \\
&&-i\frac{K_{1}}{\tau}\left(K_{1}+2K_{3}\right)\left(2M_{1}+M_{2}\right)\omega=0,
\label{dispersion_relation}
\end{eqnarray}
which has (non-trivial) solutions at $k=0$ and $k=\pi/\alpha$ (i.e., first Brillouin zone)
given by
\begin{eqnarray}
&&f_{c,1}\left(k=0\right)=
\begin{cases}
    0 \\
    i\frac{1}{2\pi\tau}
\end{cases} \quad \textrm{(lower acoustic)}, \nonumber \\
&&f_{c,2}\left(k=0\right)=\pm\frac{1}{2\pi}\sqrt{\frac{K_{1}+2K_{3}}{M_{1}}-\frac{1}{4\tau^{2}}}+i\frac{1}{4\pi\tau} \quad \textrm{(second lower optic)}, \nonumber \\
&&f_{c,3}\left(k=0\right)=\pm\frac{1}{2\pi} \sqrt{\frac{K_{1}\left(2M_{1}+M_{2}\right)}{M_{1}M_{2}}-\frac{1}{4\tau^{2}}}+i\frac{1}{4\pi\tau} \quad \textrm{(first upper optic)}, \nonumber \\
&&f_{c,4}\left(k=\pi/\alpha\right)=\pm\frac{1}{2\pi} \sqrt{\frac{K_{1}}{M_{1}}-\frac{1}{4\tau^{2}}}+i\frac{1}{4\pi\tau} \quad \textrm{(first lower optic)}, \nonumber \\
&&f_{c,5}\left(k=\pi/\alpha\right)=\pm\frac{1}{2\pi}\sqrt{\frac{K_{1}\left(2M_{1}+M_{2}\right)+2K_{3}M_{2}-\tilde{f}}{2M_{1}M_{2}}-\frac{1}{4\tau^{2}}}+i\frac{1}{4\pi\tau}  \quad \textrm{(upper acoustic)}, \nonumber \\
&&f_{c,6}\left(k=\pi/\alpha\right)=\pm\frac{1}{2\pi}\sqrt{\frac{K_{1}\left(2M_{1}+M_{2}\right)+2K_{3}M_{2}+\tilde{f}}{2M_{1}M_{2}}-\frac{1}{4\tau^{2}}}+i\frac{1}{4\pi\tau}  \quad \textrm{(second upper optic)}, \nonumber \\
\label{cut_off_freq}
\end{eqnarray}
where $\tilde{f}=\sqrt{\left[K_{1}\left(2M_{1}+M_{2}\right)+2K_{3}M_{2}\right]^{2}-16K_{1}K_{3}M_{1}M_{2}}$.
The solutions (\ref{cut_off_freq}) correspond to the \textit{cut-off frequencies}
of the spectral bands. The above values are (in principle) complex since we consider
dissipative dynamics in the system~(\ref{gc_start}) (embodied by the $\dot{u}_{n}/\tau$
term) in order to model the experimental setup. Therefore, the reported frequencies 
hereafter will correspond to the \textit{real} part of these values, i.e., the part 
contributing to the dispersion properties of the solutions rather than their decay. 
In Table~\ref{cut_off_freq_values}, the predicted values of the cut-off frequencies
are presented (up to two decimal places) using the values of the material parameters
of Table~\ref{gc_params} and $\tau = 2.1$ ms.

Finally, upon solving numerically the dispersion relation (cf. Eq.~(\ref{dispersion_relation})),
the frequency as a function of the wavenumber ($\kappa\alpha$) is presented
in Fig.~\ref{fig1b} together with the cut-off frequencies of Table~\ref{cut_off_freq_values}. 
Note that these cut-off frequencies are also plotted as horizontal dashed black
lines in Fig.~\ref{fig1a} for comparison. It can be discerned from both panels of
Fig.~\ref{fig1} that there exist two finite gaps in the frequency spectrum
(in general their number is one less than the period of the granular crystal),
namely, $[f_{c,5},f_{c,4}]$ and $[f_{c,3},f_{c,2}]$, together with one semi-%
infinite gap $[f_{c,6},\infty)$. We also note the presence of two additional
localized modes depicted in Fig.~\ref{fig1a} between the first and second 
optical pass bands due to the fixed boundary conditions.
These modes, denoted by black dots in Fig.~\ref{fig1a},
correspond to surface modes and their existence has been
previously discussed, e.g., in~\cite{Theo2010}.


\begin{table}[h]
\caption{Calculated cut-off frequencies (in $\textrm{kHz}$) using Eqs.~(\ref{cut_off_freq}).} 
\centering
\begin{tabular}{| c | c | c | c | c |}
\hline
 $f_{c,5}$  &  $f_{c,4}$ & $f_{c,3}$ & $f_{c,2}$ & $f_{c,6}$  \\
\hline\hline
   $3.04$   &  $4.23$    &  $6.04$   &  $6.71$   &  $7.37$   \\ 
\hline
\end{tabular}
\label{cut_off_freq_values}
\end{table}

\subsection{Steady-state analysis in the linear regime}
In order to account for the actuation in the linear problem we add an external
forcing term to Eq.~(\ref{gc_linear_matrix}) in the form,
\begin{equation}
\dot{\mathbf{Y}}=\mathcal{A}\,\mathbf{Y}+\mathbf{F}(\omega_{b}), 
\label{gc_linear_matrix_F}
\end{equation}
where the sole non-zero entry of $\mathbf{F}(\omega_{b})$ is at the $(N+1)$th node
and has the form $F_{N+1} = a\cos{\left(\omega_{b}t\right)}$ and the matrix 
$\mathcal{A}$ is given by Eq.~(\ref{matrix_linear_A}). Equation~(\ref{gc_linear_matrix_F})
can be solved by introducing the ansatz 
\begin{equation}
\mathbf{Y} =
\begin{pmatrix}
\begin{aligned}
a_{1}\cos{\left(\omega_{b}t\right)} &+ b_{1}\sin{\left(\omega_{b}t\right)}                      \\
&\,\,\,\vdots                                                                                         \\
a_{N}\cos{\left(\omega_{b}t\right)} &+ b_{N}\sin{\left(\omega_{b}t\right)}                      \\
-a_{1}\omega_{b}\sin{\left(\omega_{b}t\right)} &+ b_{1}\omega_{b}\cos{\left(\omega_{b}t\right)} \\
&\,\,\,\vdots                                                                                         \\
-a_{N}\omega_{b}\sin{\left(\omega_{b}t\right)} &+ b_{N}\omega_{b}\cos{\left(\omega_{b}t\right)} \\
\end{aligned}
\end{pmatrix},
\label{steady_state_lin_ansatz}
\end{equation}
with unknown (real) parameters $a_{j}$ and $b_{j}$ ($j=1,\cdots,N$). 
Inserting Eq.~(\ref{steady_state_lin_ansatz}) into Eq.~(\ref{gc_linear_matrix_F})
yields a system of linear equations which can be easily solved: 
\begin{equation}
\widetilde{\mathcal{A}} \,\mathbf{X}=\widetilde{\mathbf{F}}, 
\label{steady_state_lin_system}
\end{equation}
where
\begin{equation}
\widetilde{\mathcal{A}} =
\begin{pmatrix}
             \widetilde{\mathcal{C}} & \left(\omega/\tau\right) \mathbf{I} \\
-\left(\omega/\tau\right) \mathbf{I} & \widetilde{\mathcal{C}}
\end{pmatrix},
\label{matrix_linear_A_tilde}
\end{equation}
$\widetilde{\mathcal{C}}=-\left(\mathcal{C}+\omega_{b}^{2}\,\mathbf{I}\right)$ with 
$\mathcal{C}$ given by Eq.~(\ref{matrix_C}), while $\mathbf{X}=\left(a_{1},\dots,a_{N},b_{1},\dots,b_{N}\right)$
is the vector containing the unknown coefficients and $\widetilde{\mathbf{F}}=\left(a\,\frac{K_{0}}{M_{1}},0,\dots,0\right)$. 
We refer to Eq.~(\ref{steady_state_lin_ansatz}) as the 
asymptotic equilibrium of the linear problem (as dictated by the actuator),
since all solutions of the linear problem approach it for $t \rightarrow \infty$. 
Clearly, this state will be spatially localized if the 
forcing frequency $f_b$ lies within a spectral gap, otherwise it will be spatially
extended (with the former corresponding to a surface breather, which we discuss in 
the following section). For small driving amplitude $a$, we find that the long-term 
dynamics of the system approaches 
an asymptotic state that is reasonably well approximated 
by~\eqref{steady_state_lin_ansatz}.
However, for 
larger driving amplitudes the effect of the nonlinearity becomes significant, and we
can no longer rely on the linear analysis. 

In order to understand the dynamics in the high-amplitude regime, we perform a bifurcation
analysis of time-periodic solutions of the nonlinear problem~\eqref{gc_start} using a 
Newton-Raphson method that yields exact time-periodic solutions (to a prescribed tolerance)
and their linear stability through the computation of Floquet multipliers (see Appendix%
~\ref{sec:app_num_methods} for details). We use the asymptotic linear 
state~\eqref{steady_state_lin_ansatz}
as an initial guess for the Newton iterations.

\begin{figure}[!pt]
\begin{center}
\vspace{-0.1cm}
\mbox{\hspace{-0.2cm}
\subfigure[][]{\hspace{-0.2cm}
\includegraphics[height=.16\textheight, angle =0]{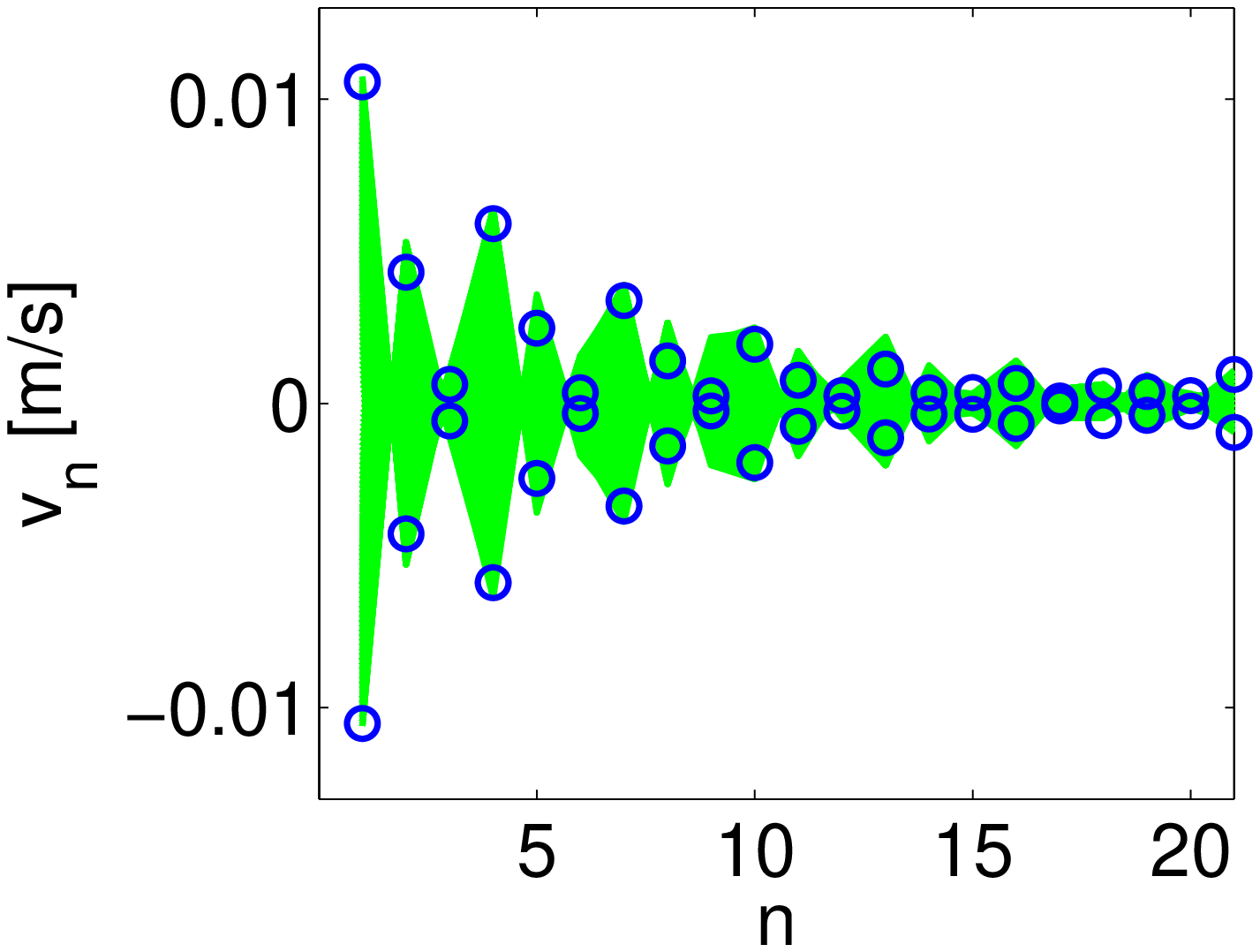}
\label{fig4a}
}
\subfigure[][]{\hspace{-0.2cm}
\includegraphics[height=.16\textheight, angle =0]{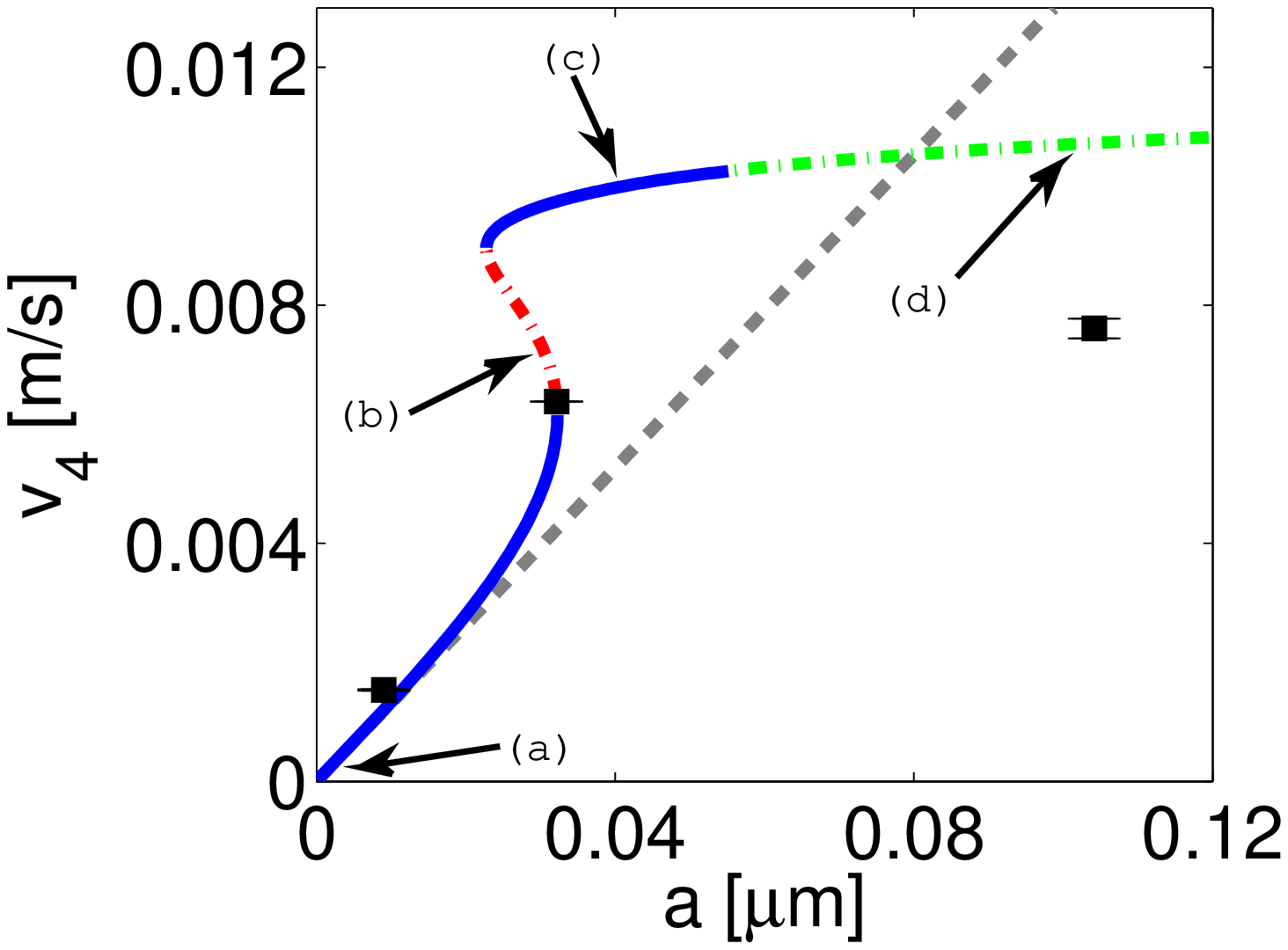}
\label{fig4b}
}
\subfigure[][]{\hspace{-0.2cm}
\includegraphics[height=.16\textheight, angle =0]{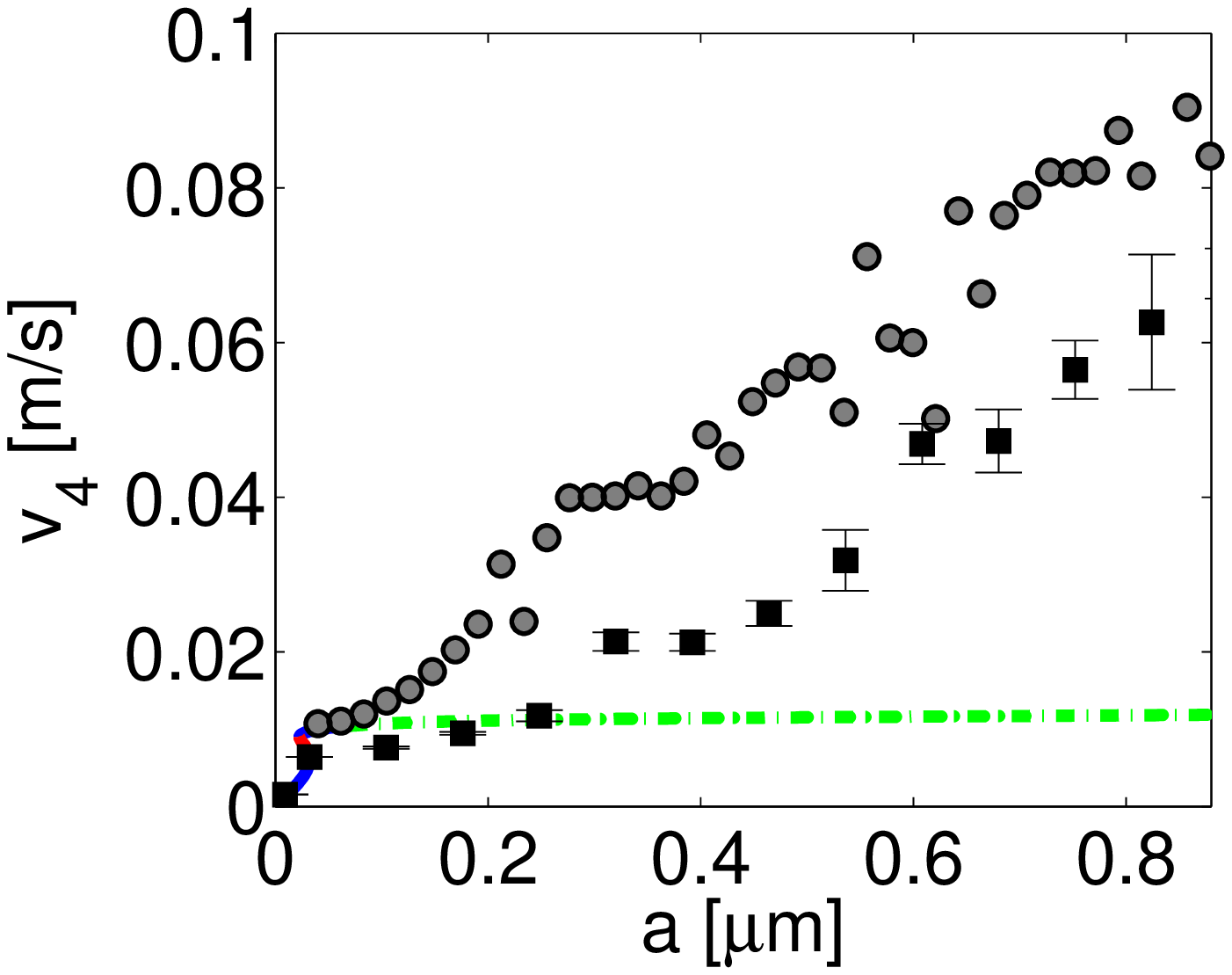}
\label{fig4c}
}
}
\end{center}
\caption{(Color online)  Comparison between experimental and theoretical results
for a driving frequency $f_{b}=6.3087\,\textrm{kHz}$, which lies in the second 
spectral gap. \textbf{(a)} Profile of solution for a driving amplitude of $a=32.16\,\textrm{nm}$. 
Experimental values (green shaded areas) are shown as a superposition of the 
velocities measured over the $50\,\textrm{ms}$ time window of recorded data. The 
numerically predicted extrema (over one period) of the exact solution obtained 
via Newton's method are shown with blue (open) circles.  \textbf{(b)} The quantity
$\varv_{4}$ 
as a function of the amplitude $a$ of the drive.  For the numerically exact solutions, 
$\varv_{4}$ is the maximum velocity of the $4$th bead over one period of motion. 
Note that the blue segments correspond to stable parametric regions while the red
and green ones correspond to real and oscillatory unstable parametric regions, 
respectively. Black squares with error bars represent the experimentally measured
mean values of the quantity $\varv_{4}$ over three experimental runs, where $\varv_{4}$
was calculated as the maximum measured over the $50\,\textrm{ms}$ time window of
recorded data. Finally, the dashed gray line is the maximum velocity (over one period)
of bead $4$ as predicted by the linear solution of Eq.~(\ref{steady_state_lin_ansatz}). 
\textbf{(c)} Zoomed out version of panel (b). The gray circles correspond to 
numerical
results that were obtained by dynamically running the equations of motion and monitoring
the maximum velocity $\varv_{4}$ over the experimental time window. The
experimental data in this panel are shown again by black squares.
}
\label{fig4}
\end{figure}

\section{Main Results} \label{main}
Besides the linear limit considered above, another relevant situation is 
that of the Hamiltonian system, i.e., with $1/\tau \rightarrow 0$ in 
Eq.~\eqref{gc_start} and $a \rightarrow 0$ in Eq.~(\ref{actuator}). 
 It is well known that time-periodic solutions that
are localized in space (i.e., breathers) exist in a host of discrete Hamiltonian
systems \cite{Flach2007}, including granular crystals with spatial periodicity
\cite{Theo2010}. In particular, Hamiltonian trimer granular chains were recently
studied in \cite{hooge13}.  There, it was 
observed that breathers with frequency
in the second spectral gap are more robust than those with frequency
in the first spectral gap. Features
that appear to enhance the stability of the higher gap breathers are (i)
tails avoiding resonances with the spectral bands and (ii)
lighter beads oscillating out-of-phase. 
In that sense, the breathers 
found in the second spectral gap of trimers are somewhat reminiscent of
breathers found in the 
gap of dimer lattices \cite{Theo2010}. Furthermore, breathers of damped-driven
dimer granular crystals (i.e., Eq.~\eqref{gc_start} with a spatial periodicity of two) 
were studied recently in \cite{hooge12}. In that setting, the breathers become \emph{surface breathers}
since they are localized at the surface, rather than the center of the chain.
Yet,
if one translates the surface breather to the center of the chain, it bears 
a strong
resemblance to a ``bulk" breather. Thus, nonlinearity, periodicity and discreteness 
enabled two classes of relevant states: The nonlinear surface breathers and ones tuned
to the external actuator (i.e., those proximal to the asymptotic linear
state dictated by the actuator). These two
waveforms were observed to collide and disappear in a limit 
cycle saddle-node bifurcation
as the actuation amplitude was increased~\cite{hooge12}. 
Beyond this critical point, no stable, periodic
solutions were found to exist in the dimer case of~\cite{hooge12}
and the dynamics were found to ``jump" to a chaotic branch. 
We aim to identify similar features in the case of the trimer with 
a particular emphasis on the 
differences arising due to the higher order periodicity of the system. 
To that end, we 
first present results on surface breathers with frequency in the second gap, 
and perform
parameter continuation in driving amplitude to draw comparisons to the 
bifurcation structure
in dimer granular lattices. Following that, we investigate breathers in the first spectral
gap and compare them to their second gap breather counterparts. 
Finally we identify both localized
and spatially extended states as the driving frequency is varied through the entire range of
spectral values covering both gaps and the three pass bands
between which they arise. 

 

\subsection{Driving in the second spectral gap} \label{gap2}
\begin{figure}
\begin{center}
\vspace{-0.1cm}
\mbox{\hspace{-0.7cm}
\subfigure[][]{\hspace{-0.2cm}
\includegraphics[height=.148\textheight, angle =0]{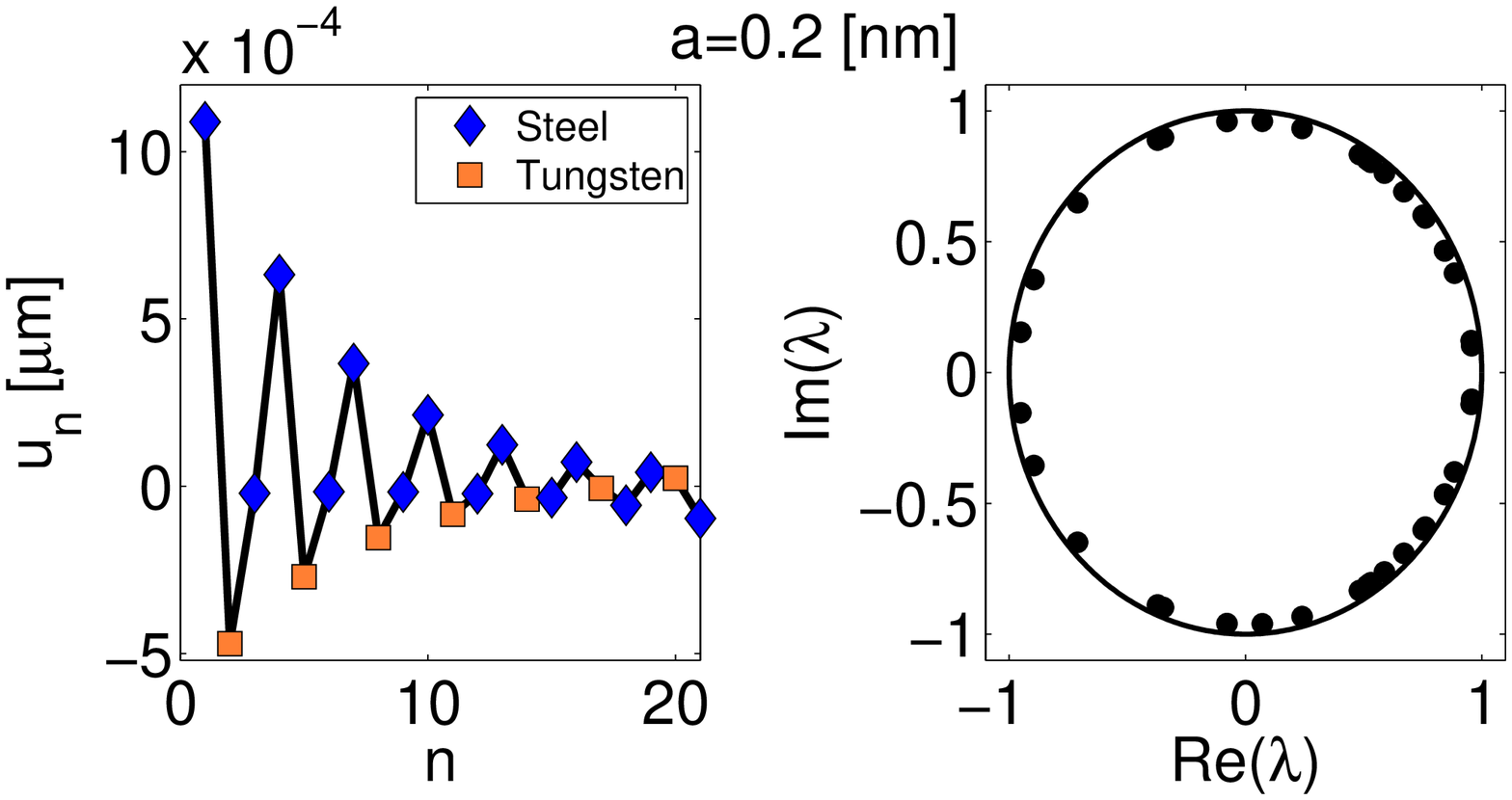}
\label{fig5a}
}
\subfigure[][]{\hspace{-0.7cm}
\includegraphics[height=.148\textheight, angle =0]{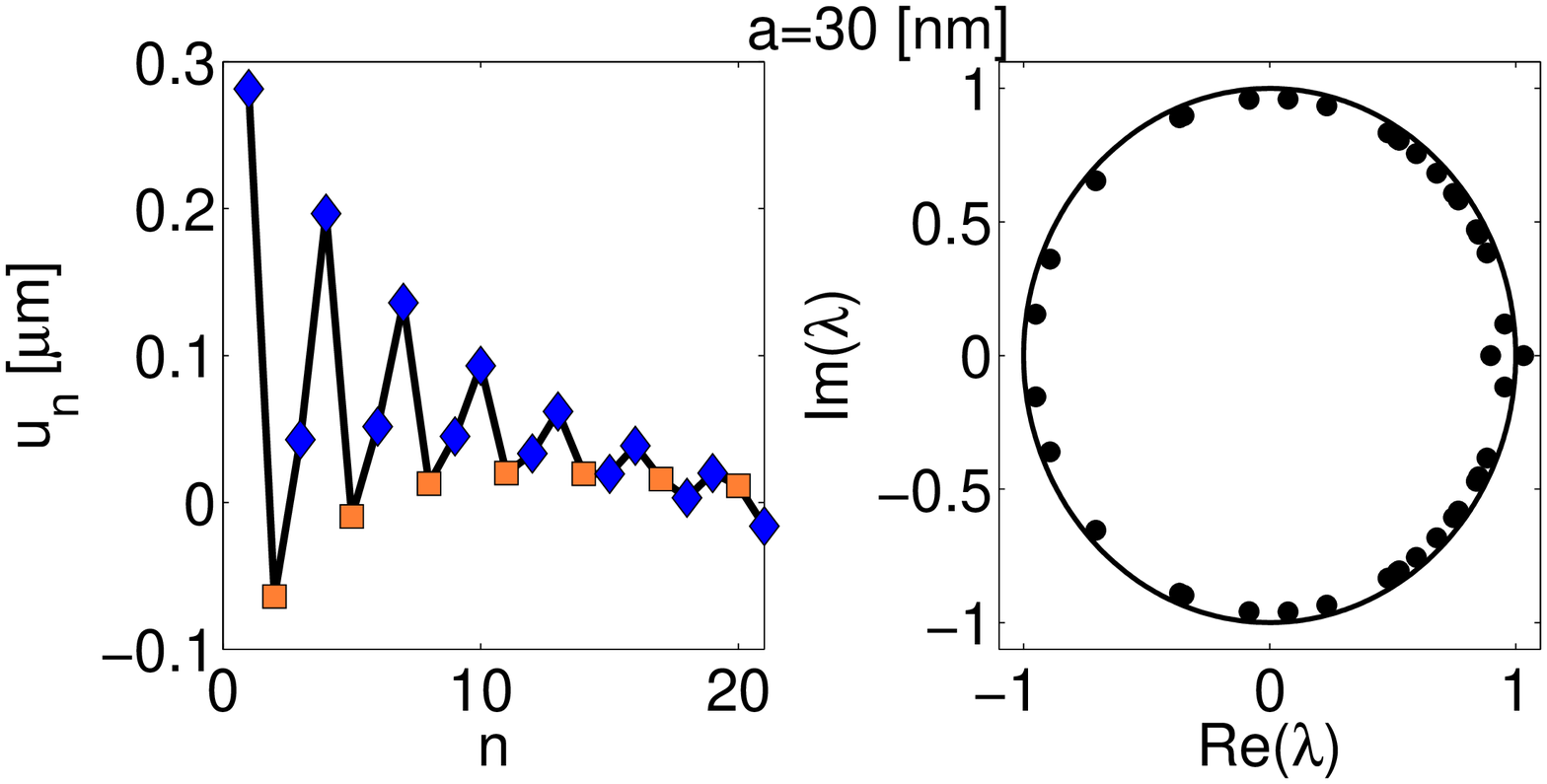}
\label{fig5b}
}
}
\mbox{\hspace{-0.7cm}
\subfigure[][]{\hspace{-0.2cm}
\includegraphics[height=.148\textheight, angle =0]{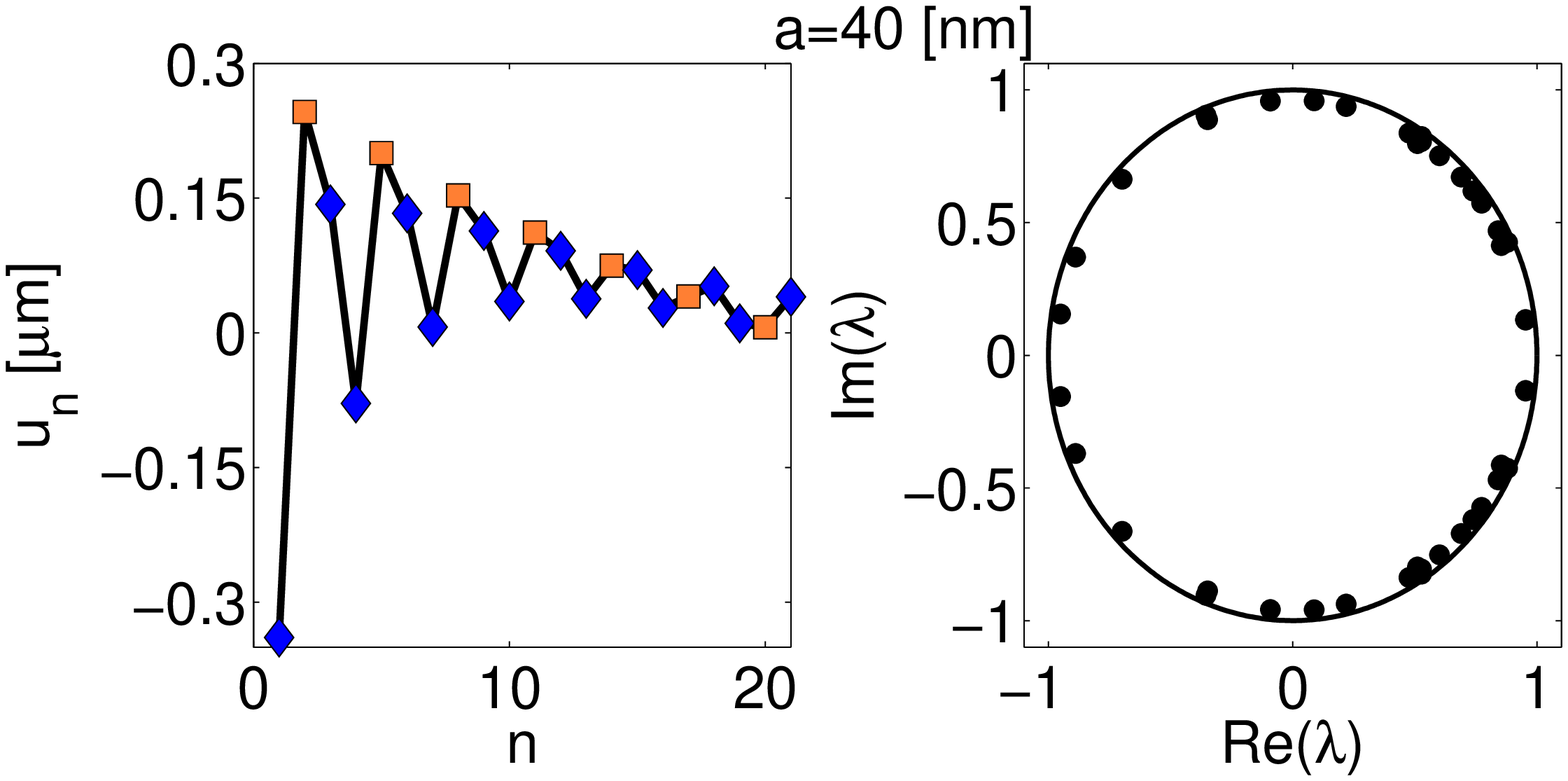}
\label{fig5c}
}
\subfigure[][]{\hspace{-0.7cm}
\includegraphics[height=.148\textheight, angle =0]{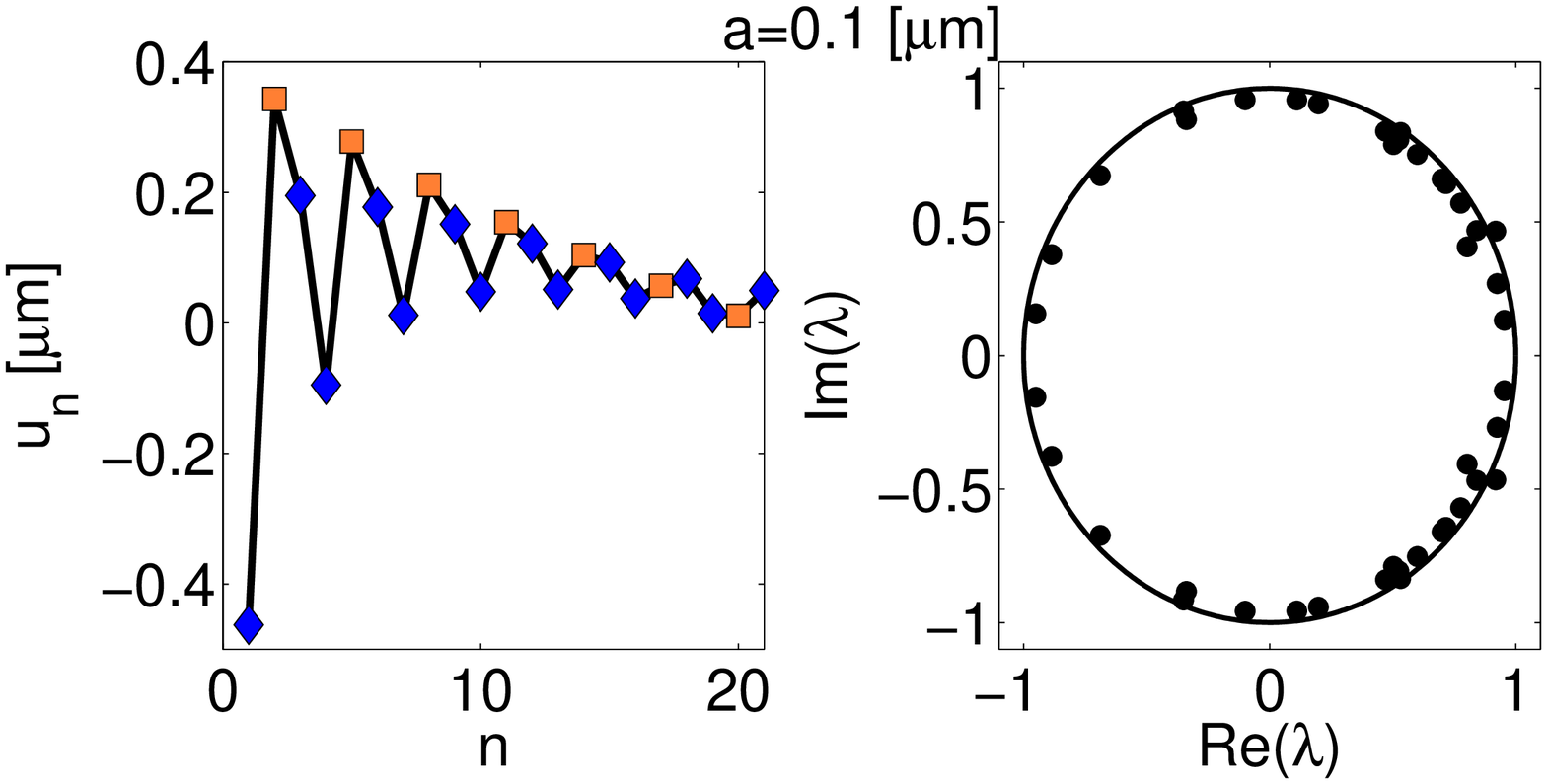}
\label{fig5d}
}
}
\end{center}
\caption{(Color online) Displacement profiles and corresponding Floquet multipliers of 
time-periodic solutions obtained at a driving frequency of $f_{b}=6.3087 \,\textrm{kHz}$
for a driving amplitude of \textbf{(a)} $a = 0.2\,\textrm{nm}$,  \textbf{(b)} $a = 30 \,\textrm{nm}$,
\textbf{(c)} $a=40\,\textrm{nm}$ and \textbf{(d)} $a = 0.1\,\textrm{$\mu$m}$. 
These solutions correspond to the (a)-(d) labels of Fig.~\ref{fig4b}. The lighter masses 
(i.e., the Steel beads) 
oscillate (nearly) out-of-phase and 
are shown as blue diamonds whereas the
heavier masses (the Tungsten Carbide ones) are shown as orange squares.
}
\label{fig5}
\end{figure}

We first consider a fixed driving frequency of $f_b = 6.3087$ kHz, 
which lies within the second spectral gap (see Table II). 
For low driving amplitude there is a single time-periodic state 
(i.e., the one proximal to the driven linear state), which the experimentally
observed dynamics follows (see panels (a) and (b) in Fig.~\ref{fig4}). 
A continuation in driving amplitude of numerically exact time-periodic
solutions (see the blue, red and green lines of Fig.~\ref{fig4b}), reveals
the existence of three branches of solutions for
a range of driving amplitudes. The branch indicated by label
(a) is proximal to the linear driven state 
given by Eq.~\eqref{steady_state_lin_ansatz}
(shown as a gray dashed line in Fig.~\ref{fig4b}). Each solution making up 
this branch is in-phase with the boundary actuator (see Fig.~\ref{fig5a}),
and has lighter masses that are out-of-phase with respect to each other 
(see e.g., Fig.~\ref{ra}). These solutions are asymptotically stable, which
is also evident in the experiments (see Fig.~\ref{fig4a} and the black markers
with error bars in Fig.~\ref{fig4b}). At a driving amplitude 
of $a\approx22.75\,\textrm{nm}$,
an unstable and stable branch of nonlinear surface breathers arise through a
saddle-node bifurcation (see labels (b) and (c) of Fig.~\ref{fig4b},
respectively). 
At the bifurcation point $a\approx22.75\,\textrm{nm}$, the profile strongly
resembles that of the corresponding Hamiltonian breather 
(when shifted to the center
of the chain), hence the name surface breather. 
It is important to note that the presence of dissipation does not
allow for this branch to bifurcate near $a \rightarrow 0$ (where this
bifurcation would occur in the absence of dissipation). Instead, the need
of the drive to overcome the dissipation ensures that the bifurcation will
emerge at a finite value of $a$.
As $a$ is increased, the solutions constituting
the unstable ``separatrix''
branch (b) resemble progressively more the ones of branch (a); 
see Fig.~\ref{fig5b}.
For example, branch (b) progressively becomes in-phase with the actuator as $a$ is increased. Indeed 
these two branches
collide and annihilate at $a\approx32.26\,\textrm{nm}$. On 
the other hand, the
(stable, at least for a parametric interval in $a$) 
nonlinear surface breather (c) 
is out-of-phase with the actuator (see Fig.~\ref{fig5c}). 
This solution loses its stability through a Neimark-Sacker bifurcation, 
which is the 
result of a Floquet multiplier (lying off the real line) 
acquiring modulus greater than
unity, (see label (d) of Fig.~\ref{fig4b} and Fig.~\ref{fig5d}). Such a Floquet multiplier
indicates concurrent growth and oscillatory dynamics of 
perturbations and thus the 
instability is
deemed as an oscillatory one. Solutions with an oscillatory instability 
are marked in 
green in Fig.~\ref{fig4b}, whereas red dashed lines correspond to 
purely real instabilities
(see also Fig.~\ref{fig5b} as a case example) and 
solid blue lines denote asymptotically 
stable regions. The reason for making the distinction between real and oscillatory instabilities
is that quasi-periodicity and chaos often lurk in regimes in 
parameter space where solutions
possess such instabilities~\cite{hooge12}. Indeed, 
past the above saddle-node bifurcation, as the amplitude
is further increased, for an additional narrow parametric
regime, the computational 
dynamics (gray circles in Fig.~\ref{fig4c}) follows the upper
nonlinear surface mode of branch (c); yet, once this branch
becomes unstable the dynamics appears to reach a chaotic state.
In the experimental dynamics (black squares in Fig.~\ref{fig4c}),
a very similar pattern is observed qualitatively, 
although the quantitative details
appear to somewhat differ. 
Admittedly, as the more nonlinear regimes of the
system's dynamics are accessed (as $a$ increases), such disparities are progressively
more likely 
due to the opening of gaps between the spheres
and the 
limited applicability (in such regimes) of the simple
Hertzian contact law.



\subsection{Driving in the first spectral gap}
\begin{figure}[!pt]
\begin{center}
\vspace{-0.1cm}
\mbox{\hspace{-0.2cm}
\subfigure[][]{\hspace{-0.2cm}
\includegraphics[height=.16\textheight, angle =0]{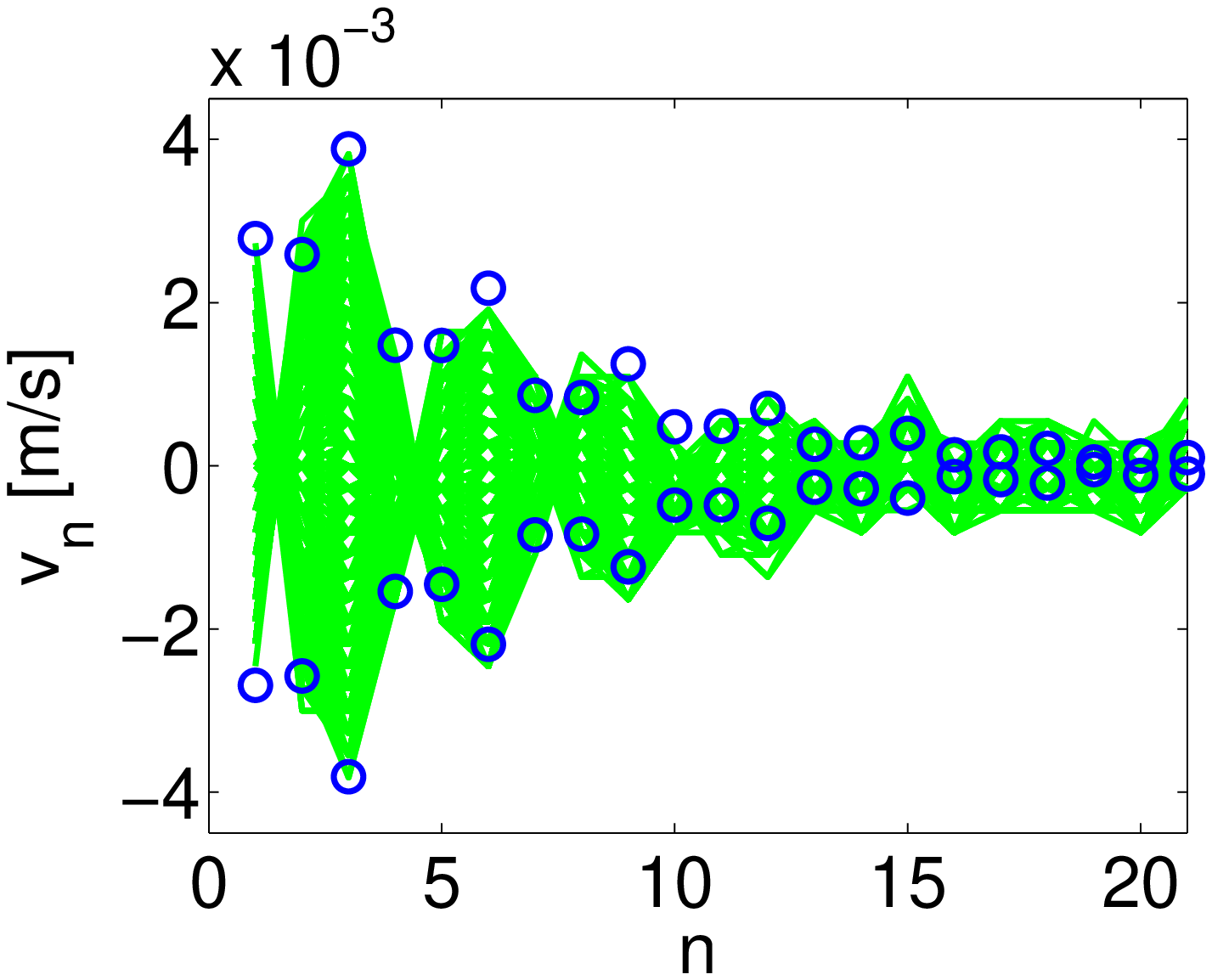}
\label{fig2a}
}
\subfigure[][]{\hspace{-0.2cm}
\includegraphics[height=.16\textheight, angle =0]{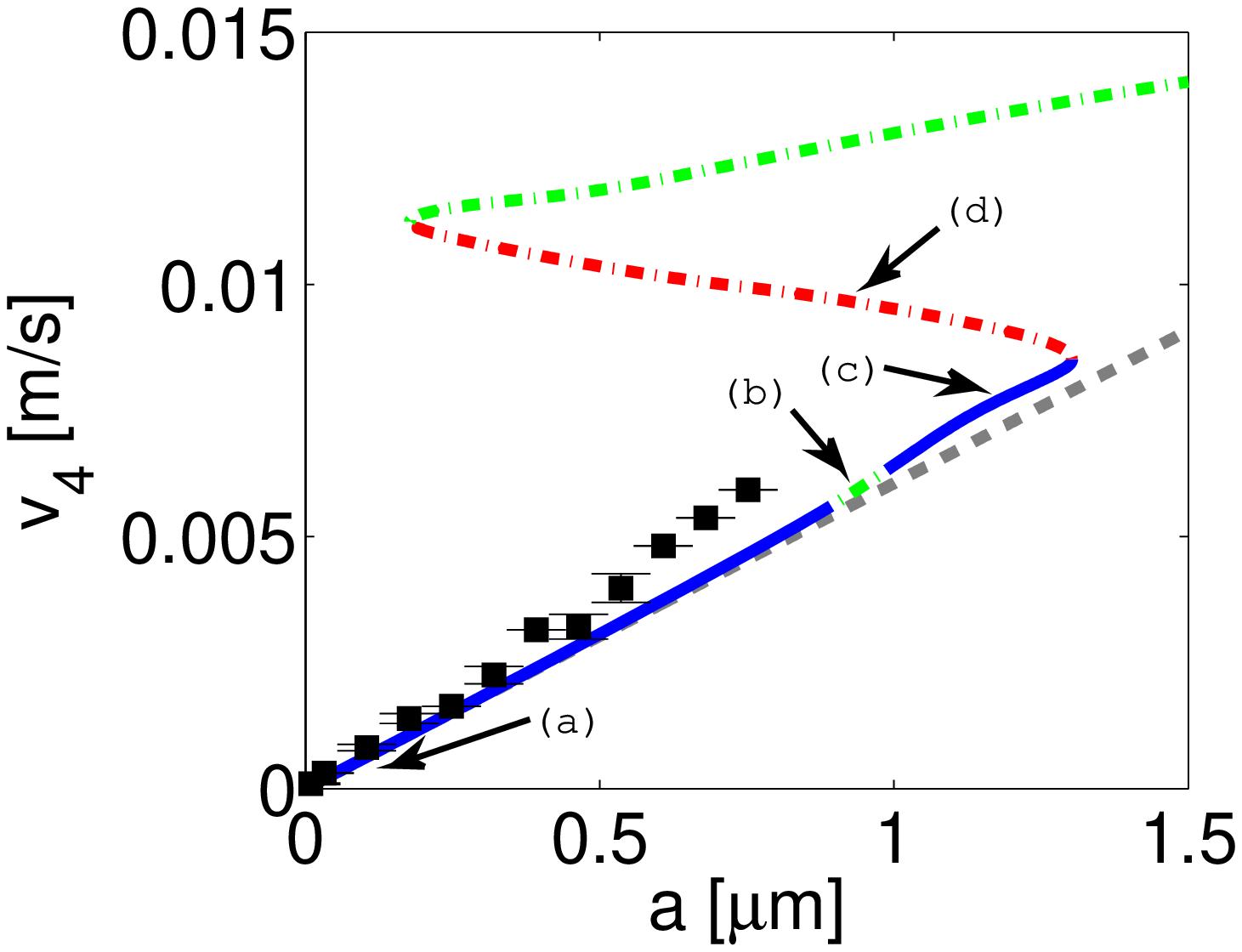}
\label{fig2c}
}
\subfigure[][]{\hspace{-0.2cm}
\includegraphics[height=.16\textheight, angle =0]{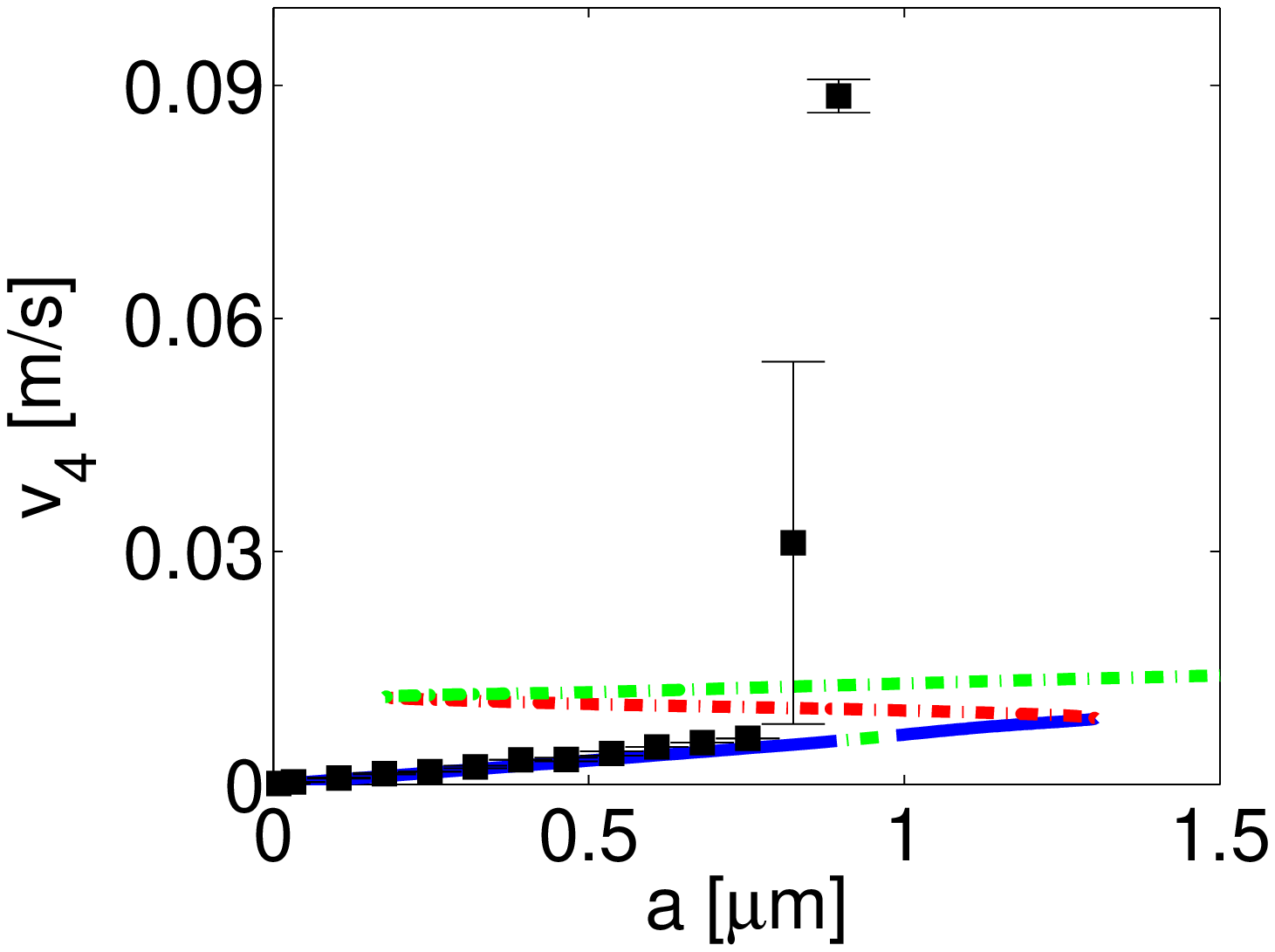}
\label{fig2b}
}
}
\end{center}
\caption{ (Color online) 
Same as Fig.~\ref{fig4} but for a driving frequency of $f_{b}= 3.7718\,\textrm{kHz}$, 
which lies in the first spectral gap. \textbf{(a)} Profile of solution for a driving 
amplitude of $a=0.24816\,\textrm{$\mu$m}$. \textbf{(b)} The bifurcation diagram for
$f_{b}= 3.7718\,\textrm{kHz}$ (shown here) is similar to the one presented in 
Fig.~\ref{fig4b}, which was for a driving frequency in the second gap ($f_{b}=6.3087\,\textrm{kHz}$).
However, the bifurcation points occur for much larger values of the driving amplitude
and the structure of the solutions themselves varies 
considerably in comparison to the second gap (see text). 
%
\textbf{(c)} Zoomed out version of panel (b). In this case, the jump of the experimental 
data is likely due to driving out of a controllable range, rather than to chaos, as in the
case of the second gap breathers. 
}
\label{fig2}
\end{figure}

\begin{figure}[!pt]
\begin{center}
\vspace{-0.2cm}
\mbox{\hspace{-0.7cm}
\subfigure[][]{\hspace{-0.2cm}
\includegraphics[height=.148\textheight, angle =0]{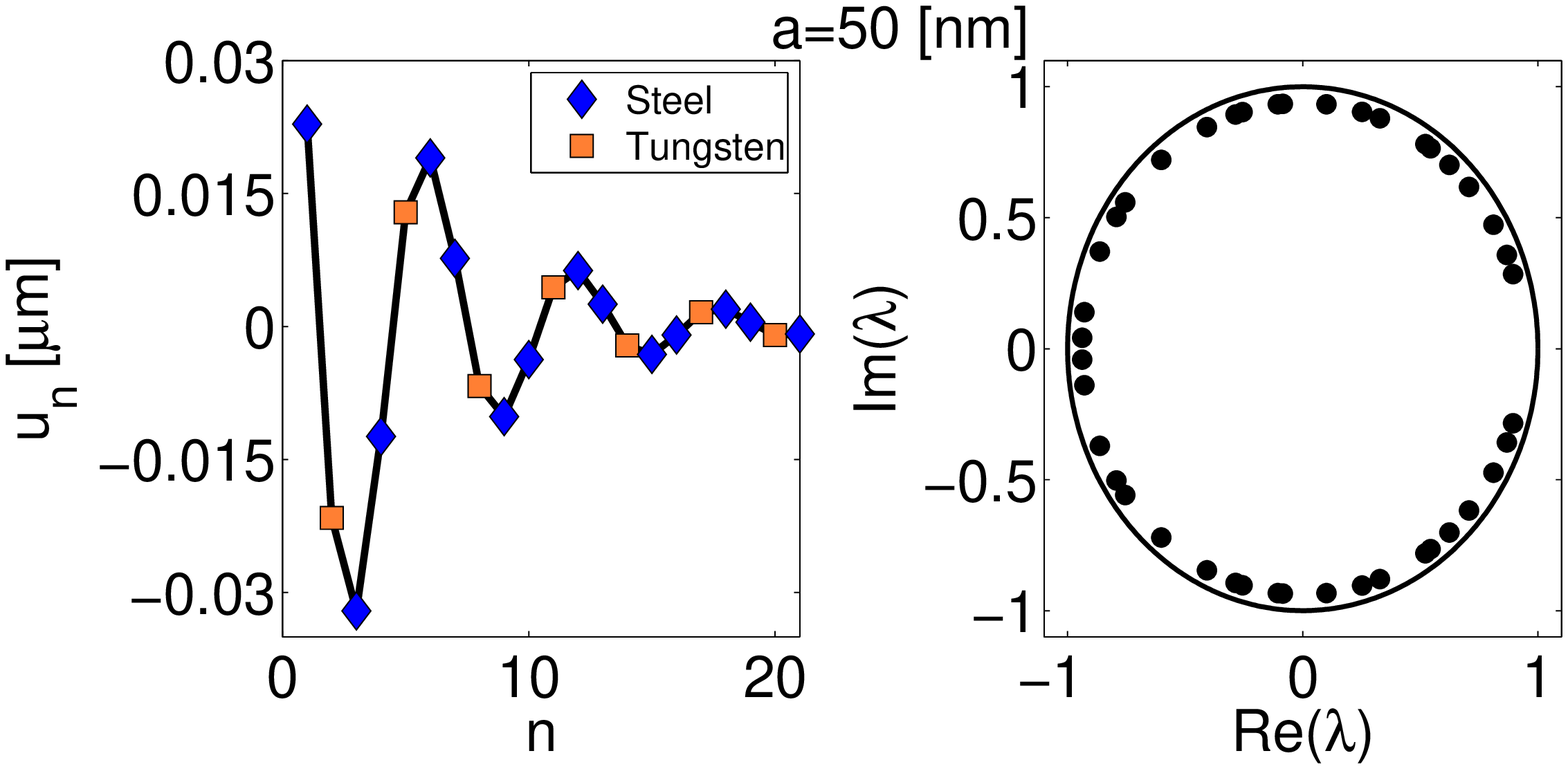}
\label{fig3a}
}
\subfigure[][]{\hspace{-0.7cm}
\includegraphics[height=.148\textheight, angle =0]{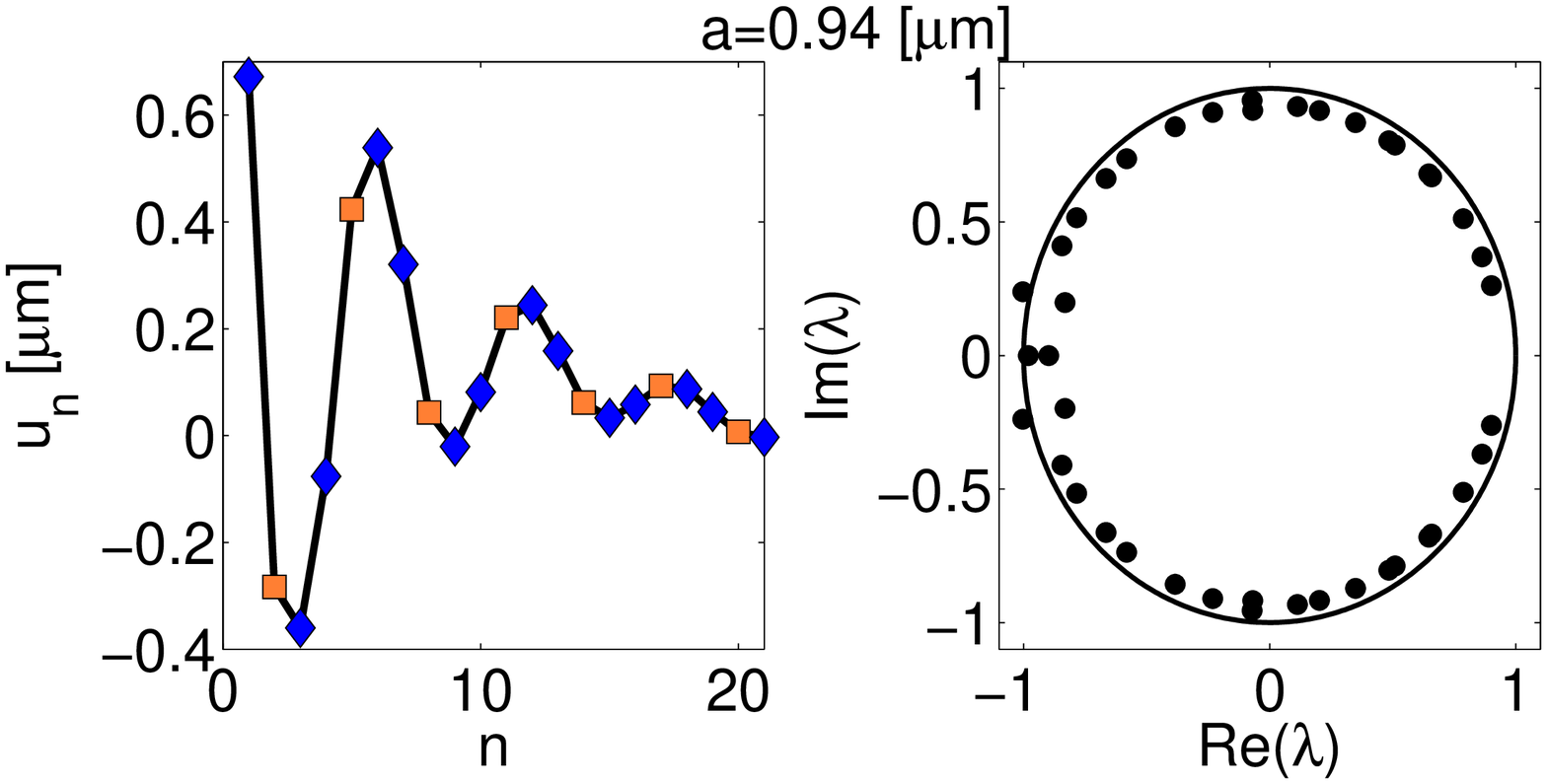}
\label{fig3b}
}
}
\mbox{\hspace{-0.7cm}
\subfigure[][]{\hspace{-0.2cm}
\includegraphics[height=.148\textheight, angle =0]{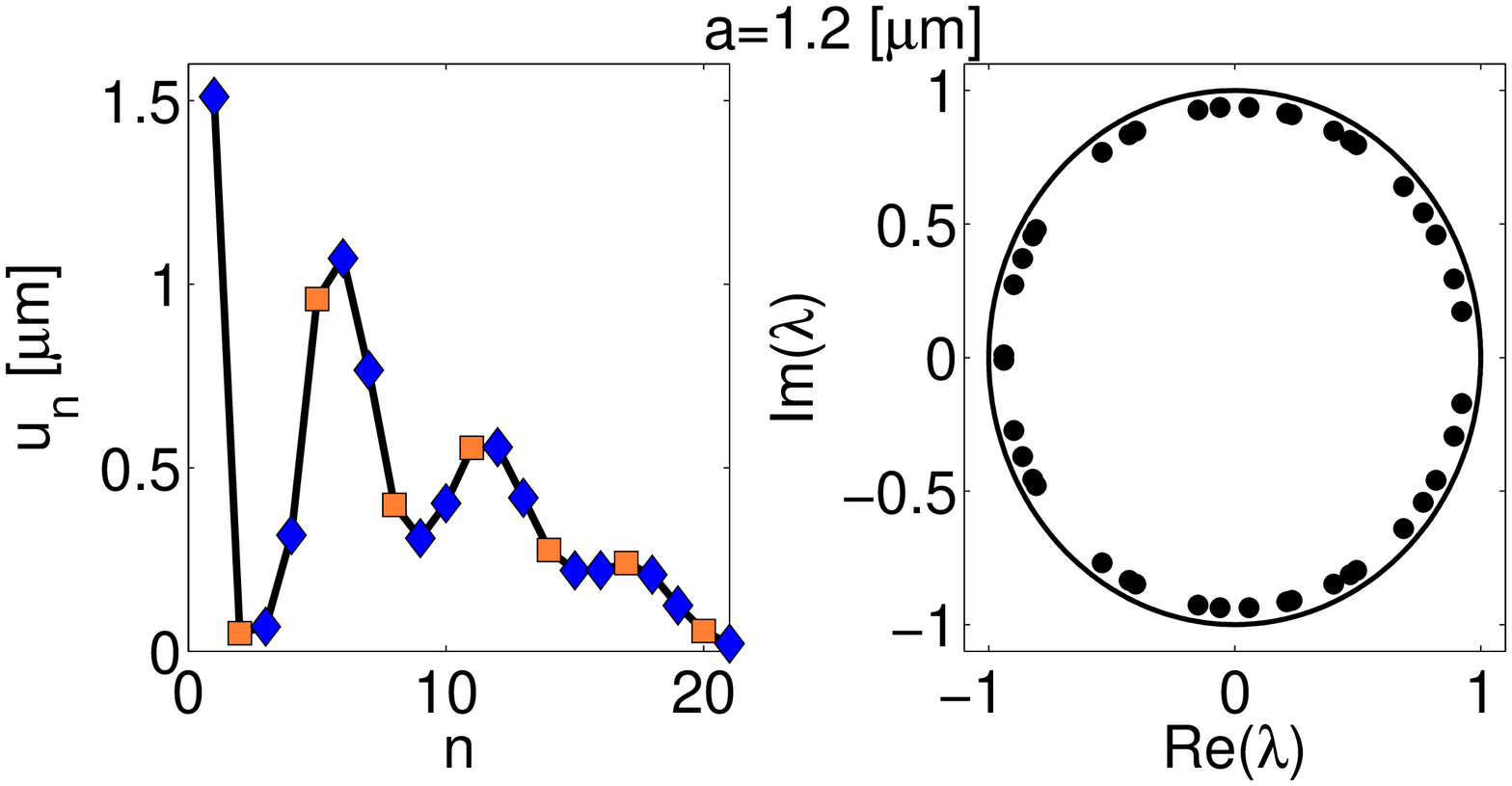}
\label{fig3c}
}
\subfigure[][]{\hspace{-0.7cm}
\includegraphics[height=.148\textheight, angle =0]{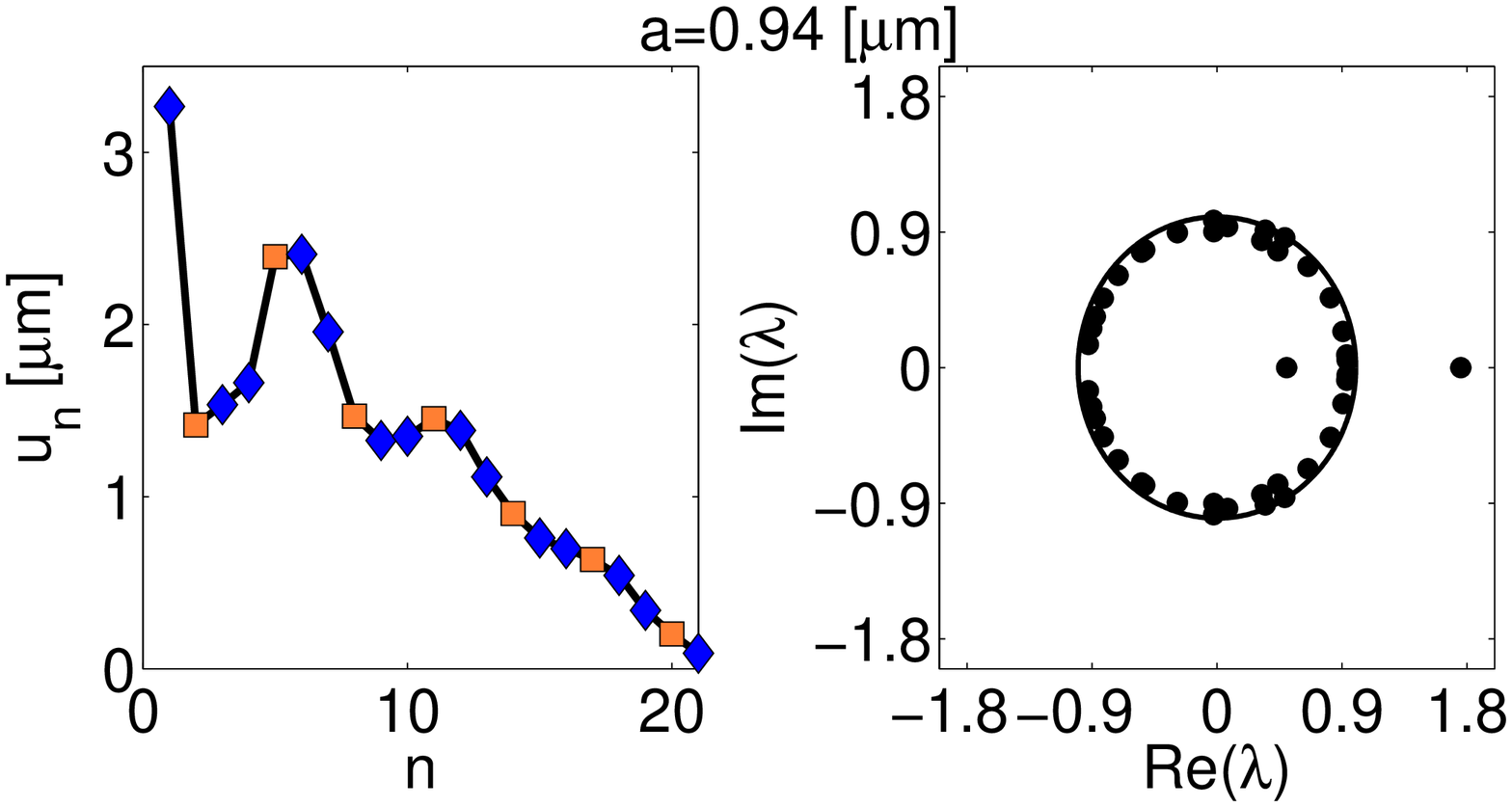}
\label{fig3d}
}
}
\end{center}
\caption{(Color online) 
Displacement profiles and corresponding Floquet multipliers of time-periodic solutions obtained
at a driving frequency of $f_{b}= 3.7718\,\textrm{kHz}$ for a driving amplitude of \textbf{(a)} 
$a=50 \,\textrm{nm}$, \textbf{(b)} $a = 0.94 \,\textrm{$\mu$m}$, \textbf{(c)} 
$a = 1.2\, \textrm{$\mu$m}$ and \textbf{(d)} $a = 0.94\,\textrm{$\mu$m}$. These solutions 
correspond to the (a)-(d) labels of Fig.~\ref{fig2c}. The lighter masses (i.e., the Steel beads)
oscillate in-phase and are shown as blue diamonds whereas the heavier masses (the Tungsten Carbide
ones) are shown as orange squares.
}
\label{fig3}
\end{figure}

\begin{figure}[!pt]
\begin{center}
\mbox{\hspace{-0.2cm}
\subfigure[][]{\hspace{-0.05cm}
\includegraphics[height=.15\textheight, angle =0]{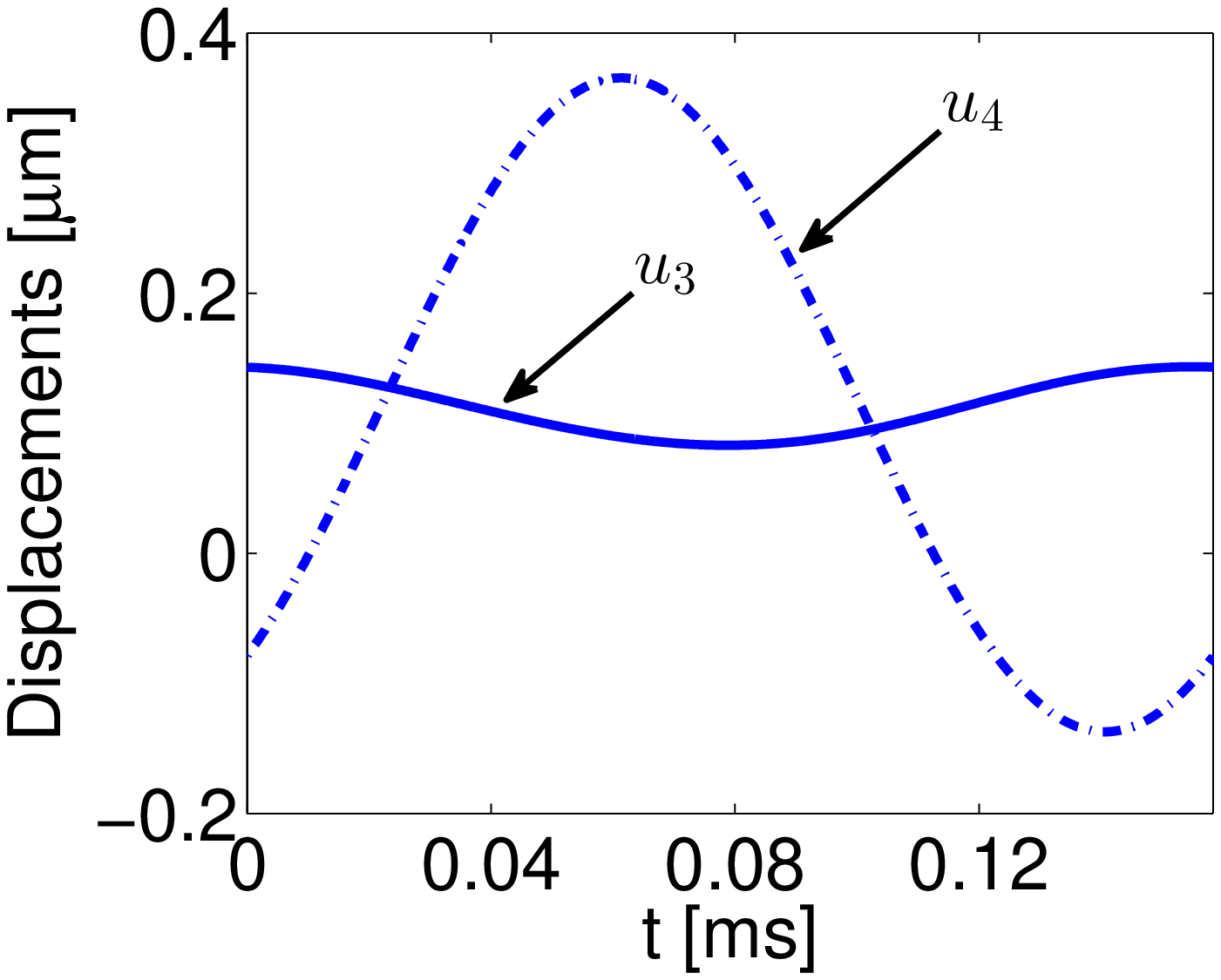} \label{ra}
}
\subfigure[][]{\hspace{-0.05cm}
\includegraphics[height=.15\textheight, angle =0]{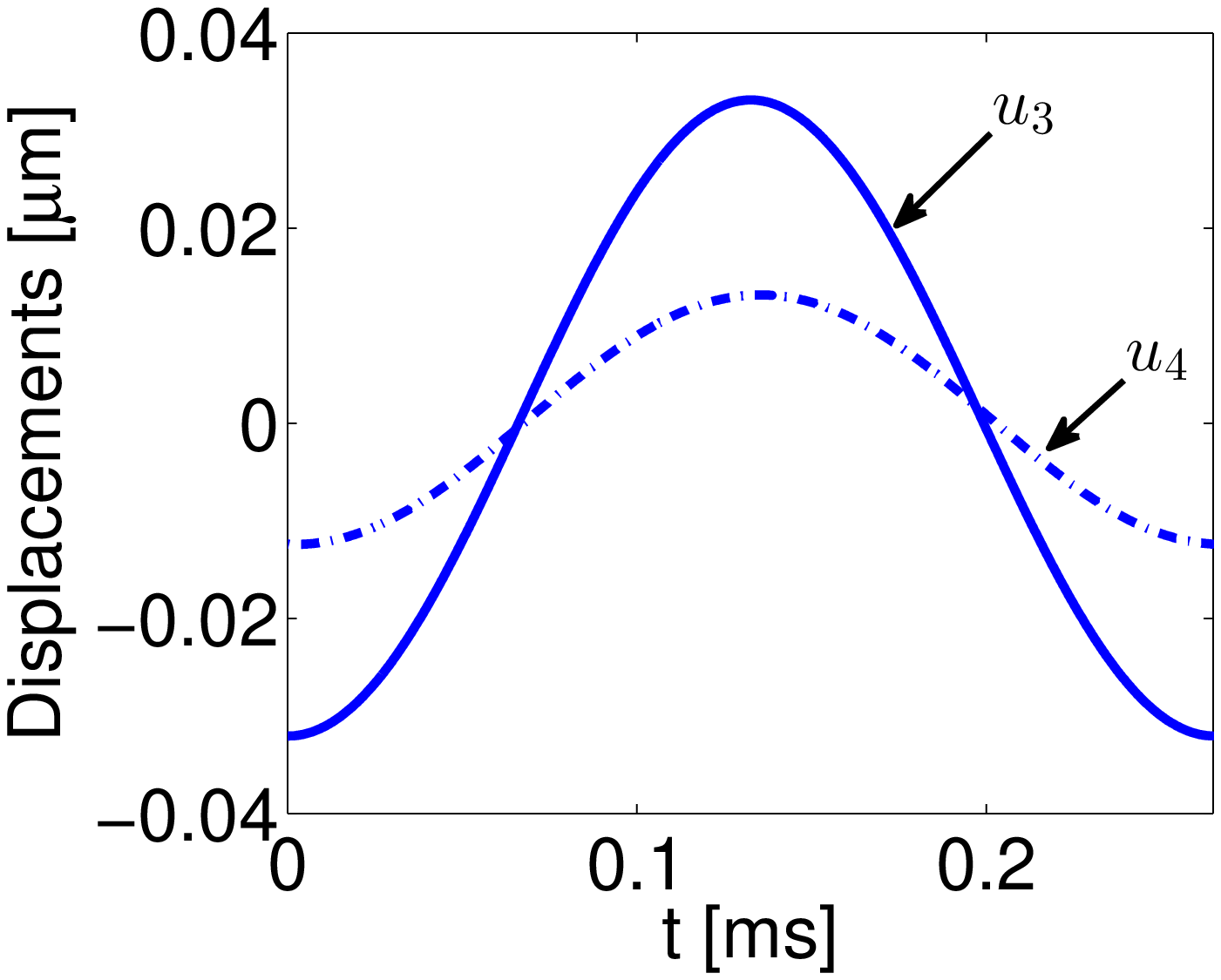} \label{rb}
}
\subfigure[][]{\hspace{-0.05cm}
\includegraphics[height=.15\textheight, angle =0]{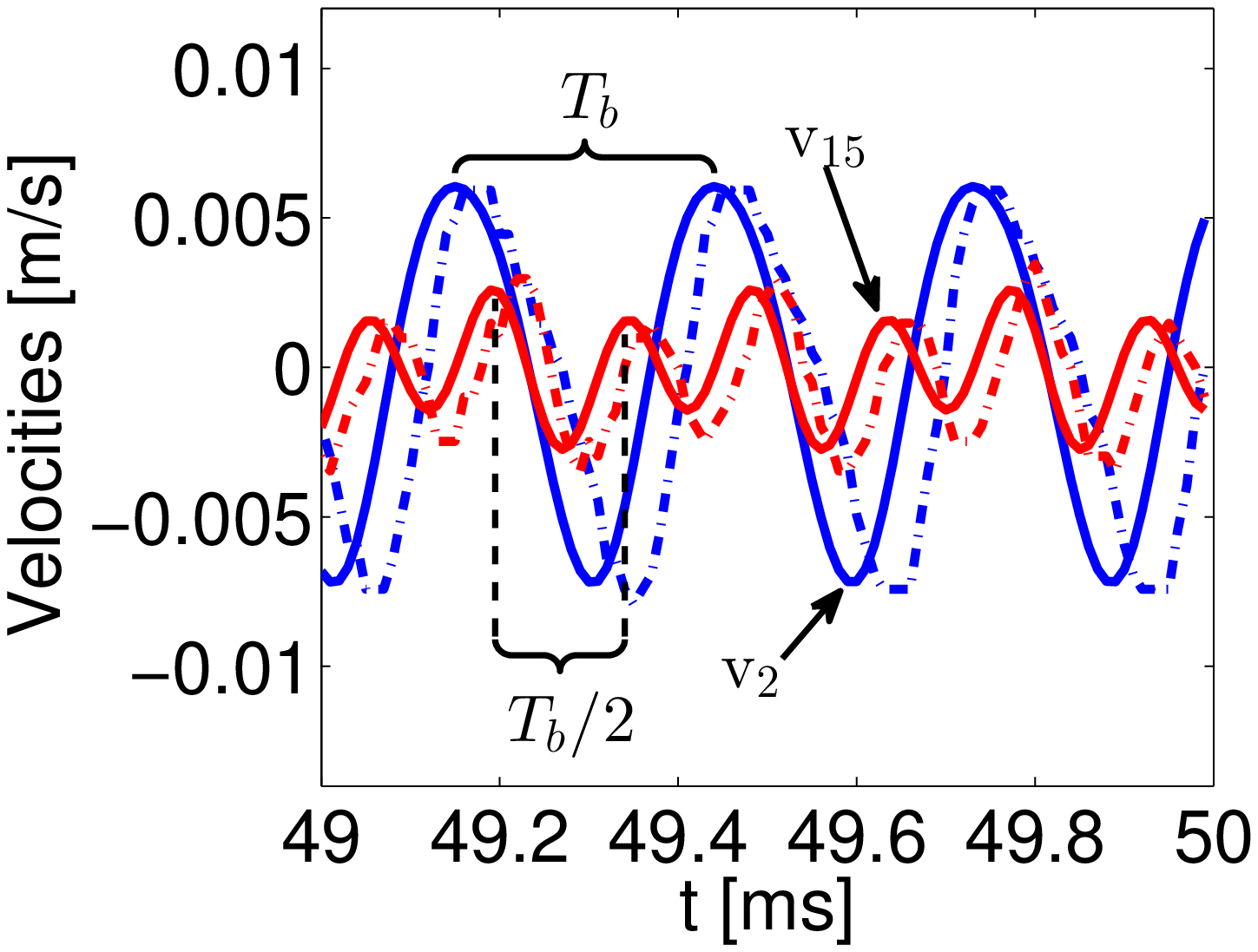} \label{rc}
}
}
\end{center}
\caption{(Color online) \textbf{(a)} Displacement vs. time over one period of motion
of two adjacent steel masses $u_3(t)$ and $u_4(t)$ for a solution in the second spectral
gap (with $f_{b}=6.3087\,\textrm{kHz}$ and $a=40 \,\textrm{nm}$). In this case the motion is (nearly)
out-of-phase. \textbf{(b)} Same as (a), but for a solution in the first spectral gap (with  
$f_{b}=3.7718\,\textrm{kHz}$ and $a=50 \,\textrm{nm}$), in which case the motion is in-phase. 
\textbf{(c)} Velocities vs. time for the motion of steel masses $\varv_{2}(t)$ and $\varv_{15}(t)$
for a solution with frequency $f_{b}=3.4429\,\textrm{kHz}$ and amplitude $a=50 \,\textrm{nm}$. 
Solid lines correspond to numerical results while the dash-dotted lines correspond to experimental data 
(mean of the three runs). Notice that 
the tail oscillates with twice the frequency of the primary nodes near the 
left boundary, where $2 f_b$ lies in the second optical band.
}
\label{resonate}
\end{figure}

We now consider a fixed driving frequency of $f_b = 3.7718$ kHz, which lies
within the first spectral gap. Qualitatively, the breather solutions and their
bifurcation structure are similar to those in the second gap (compare Figs.~\ref{fig4}
and~\ref{fig5}  with Figs.~\ref{fig2} and ~\ref{fig3}). However, there are 
several differences, which we highlight here. For example, the lighter masses
now oscillate (nearly) in-phase, rather that out-of-phase (for comparison, see panels (a) and (b) in Fig.~\ref{resonate}). 
The emergence of the nonlinear surface modes occurs for a much larger value of
the driving amplitude ($a \approx 0.1778\,\textrm{$\mu$m}$) as well as the bifurcation
of the state (a) dictated by the actuator
with the relevant branch of 
the nonlinear surface modes ($a \approx 1.3\,\textrm{$\mu$m}$). Indeed, the
latter turns out to be out-of-range for experimentally controllable driving 
amplitudes given the stroke of the piezoelectric actuator and the power 
by the electric amplifier in our experimental setup.
Thus, applications such as bifurcation based rectification \cite{Nature11}
are not suitable for breather frequencies in the first gap in the
present setting. Although the range
of amplitudes yielding stable solutions is much larger, these solutions are 
still
effectively linear (see the dashed gray line in Fig.~\ref{fig2c}). Another peculiar
feature particular to this case 
is that the branch (a)
loses its stability through a Neimark-Sacker bifurcation ($a\approx 0.9\,\textrm{$\mu$m}$),
rather than through a saddle-node bifurcation with the nonlinear surface mode, 
although it regains stability shortly thereafter ($a\approx 0.9876\,\textrm{$\mu$m}$). 
Interestingly, however, the corresponding experimental branch deviates
from the theoretical (near-linear) branch close to this destabilization
point, apparently leading to large amplitude, chaotic behavior thereafter.

As an additional feature worth mentioning, 
there are breathers within the first gap that resonate with the
linear modes. For example, any breather with a frequency $f_b \in [ 3.3550, 3.6850]$
will have a second harmonic that lies in the second optical band. Indeed, the breather
solution depicted in Fig.~\ref{rc} has a frequency $f_b=3.4429$, but the tail 
oscillates with twice that frequency (i.e., it is resonating with the linear mode
at a frequency of $2f_b$). Finally, we note that the magnitude of the instabilities
tends to be much larger in the first spectral gap, due possibly to their spatial
structure (see e.g., Fig.~\ref{fig3d}), in agreement with what was 
found in~\cite{hooge13}.

While a detailed probing of the spatial profiles and the corresponding 
bifurcation 
structure in the first and second gap reveal several differences, 
a common theme 
is that the experimental data points generally follow
rather accurately the theoretically 
predicted
curve. In particular, the agreement between experiment and theory is 
rather satisfactory
as long as the data points are within the stable parametric regions. 
This is 
clearly depicted in panel (b) of Figs.~\ref{fig4} and \ref{fig2}. Note that
the 
linear asymptotic equilibrium (shown by the 
dashed gray line) coincides with the theoretically
predicted curve as soon as Eq.~(\ref{lin_limit}) holds (a feature
that we also use as a consistency check 
for the numerics). In both cases (first or second gap) once the amplitude 
$a$ of 
the actuator reaches a critical value 
($a\approx0.32016\,\textrm{$\mu$m}$ for the second
gap and $a\approx0.82416\,\textrm{$\mu$m}$ for the first gap), the 
experimental data 
experience jumps, possibly leading to chaotic behavior. 
Importantly, these jumps arise close to the destabilization
points of the respective branches.
Nevertheless, it should also be pointed out that in some of these
regimes (especially so in the case of Fig.~\ref{fig2}), 
these driving amplitudes may be near or beyond the regime
of (accurately) controllable experimentally accessible amplitudes.


\begin{figure}[!pt]
\begin{center}
\vspace{-0.2cm}
\mbox{\hspace{-0.7cm}
\subfigure[][]{\hspace{-0.2cm}
\includegraphics[height=.32\textheight, angle =0]{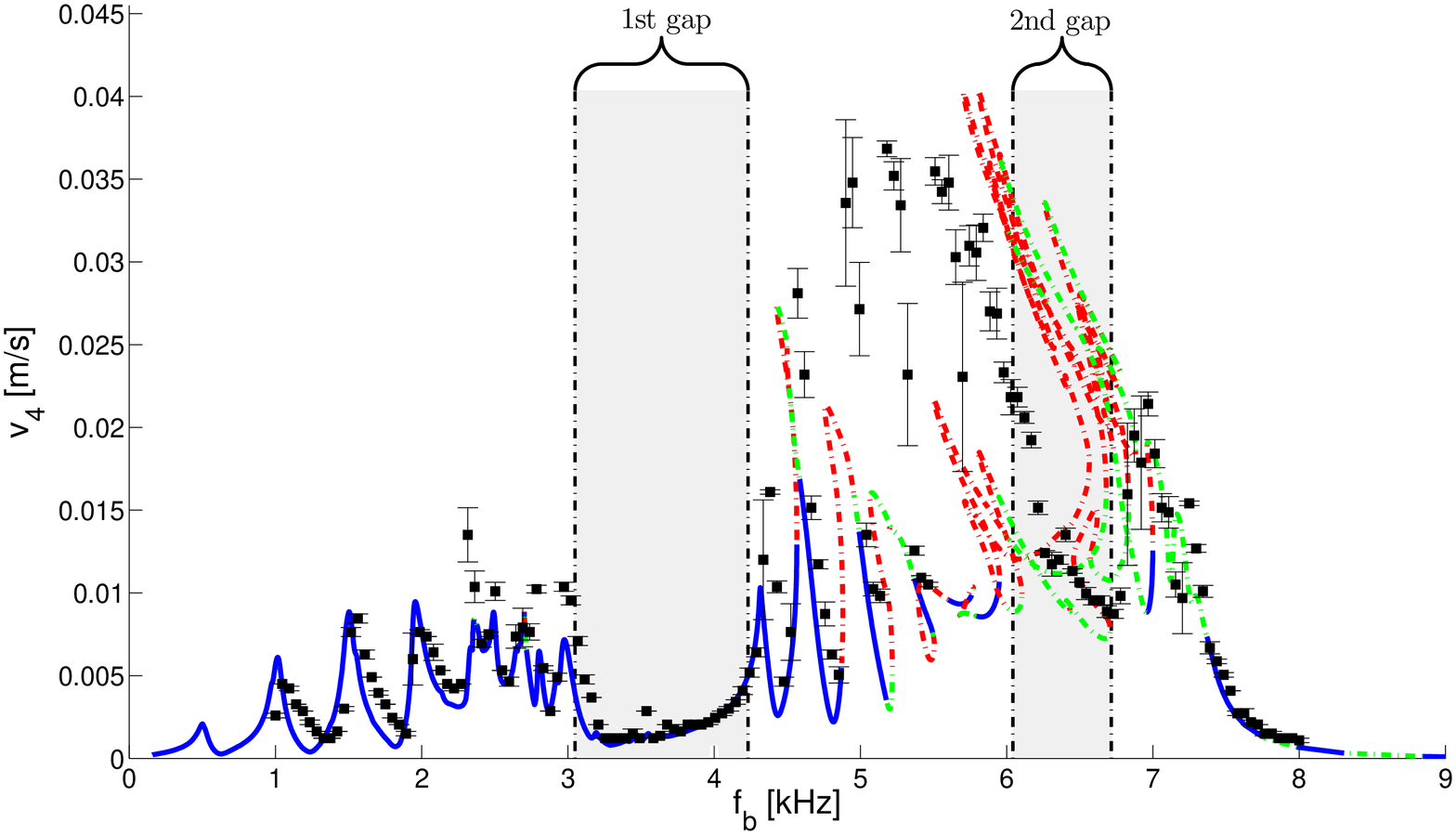}
\label{fig6a}
}
}
\mbox{\hspace{-0.7cm}
\subfigure[][]{\hspace{-0.2cm}
\includegraphics[height=.32\textheight, angle =0]{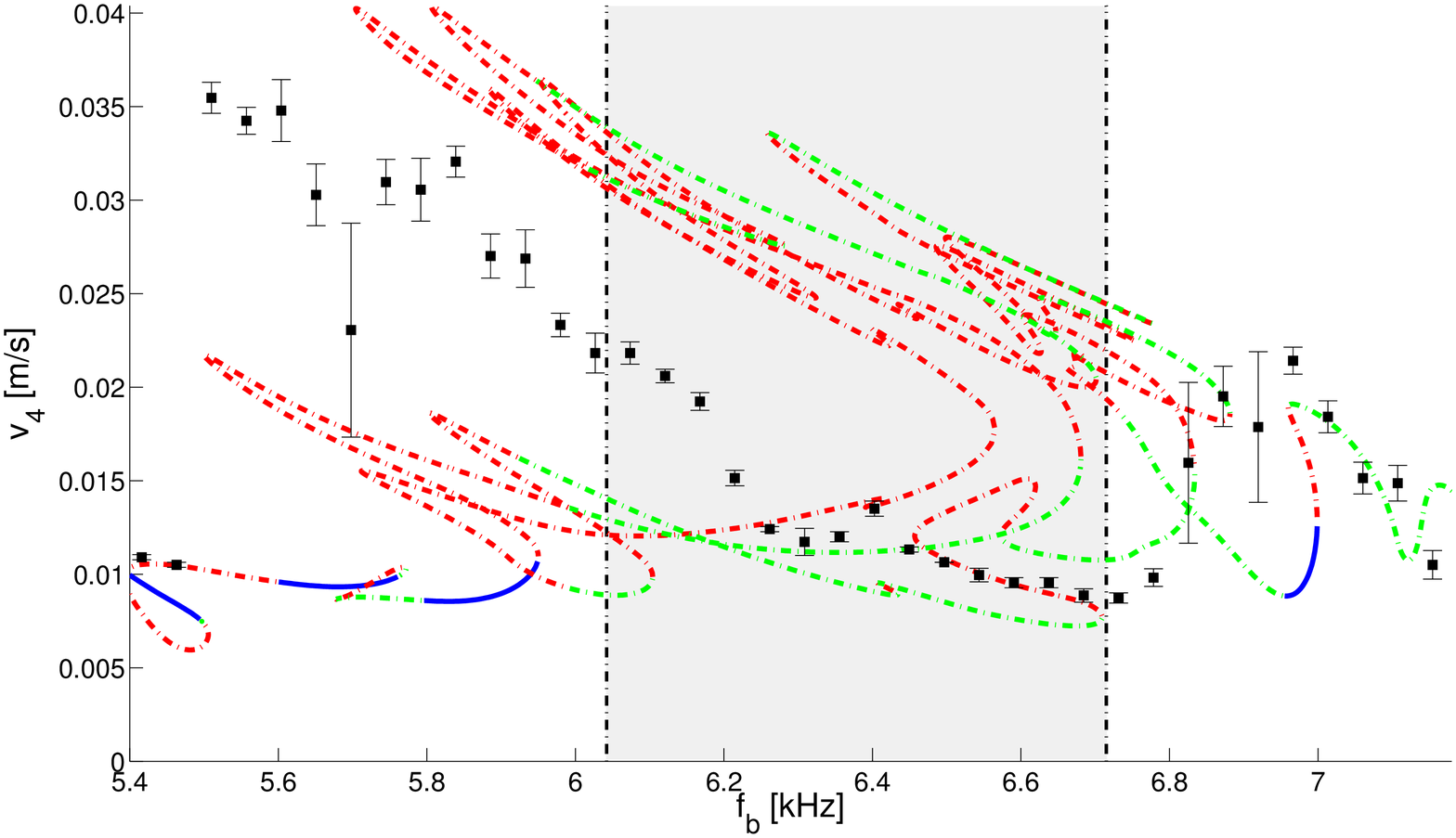}
\label{fig6b}
}
}
\end{center}
\caption{(Color online) The bifurcation diagram corresponding to the maximum velocity
of bead $4$ as a function of the actuation frequency $f_{b}$ and for a value of the 
actuation amplitude of $a=0.24816\,\textrm{$\mu$m}$ is presented with smooth curves.
The color indexing is the same as in Figs.~\ref{fig4} and \ref{fig2}.
The black squares with error bars represent the experimentally measured mean values of the 
velocities which were obtained using three experimental runs. Furthermore, the light
gray areas enclosed by vertical dash-dotted black lines correspond to the frequency
gaps according to Table~\ref{cut_off_freq_values} (see also, Fig.~\ref{fig1}). Note
that panel (b) is a zoom-in of panel (a).}
\label{fig6}
\end{figure}

\subsection{A full driving frequency continuation}
Our previous considerations suggest that the driving frequency plays an
important role in the observed dynamics of the time-periodic solutions. 
Figure~\ref{fig6} complements those results by showing a frequency
continuation for a fixed driving amplitude $a=0.24816\,\textrm{$\mu$m}$. 
For this driving amplitude, the response is highly nonlinear, where the
ratio of dynamic to static compression ranges as high as e.g. $|u_n - u_{n+1}|/ \delta_{0,n} =  1.5$. 
In the low amplitude (i.e., near linear) range, the peaks in the
force response
coincide with the eigenfrequency spectrum presented in Fig.~\ref{fig1a}, 
with all solutions being stable. For the chosen value of driving amplitude,
the features of the diagram at a low driving frequency are
similar to their linear counterparts (see e.g. the smooth resonant peaks
in the acoustic band of Fig.~\ref{fig6a}). Similar to Figs.~\ref{fig4} and
\ref{fig2}, the experimentally obtained values match the 
numerically obtained, dynamically stable response
quite well, with a few notable exceptions: (i) in narrow parametric
regions around $f_{b}\approx 2.35\,\textrm{kHz}$ and $f_{b}\approx 2.7\,\textrm{kHz}$
where the experimental points experience jumps and (ii) the experimentally
observed velocities for the same frequency 
are upshifted when compared with the theoretical curve
(see e.g. the region $f_{b}\approx 1-5.2 \,\textrm{kHz}$ in Fig.~\ref{fig6a}
where the peaks do not align exactly). Possible explanations for such 
discrepancies are given below, but we also note that discrepancies in 
experimental and theoretical dispersion curves of granular crystals  
have also been previously discussed; see, e.g., for a recent 
example~\cite{Tun_Band_gaps}.

In general, as the driving amplitude is increased, the peaks in the bifurcation
diagram bend (in some cases, they bend 
into the gap) in a way reminiscent of the familiar 
fold-over event of the driven oscillators; see, e.g.,~\cite{tristram} for
a recent study extending this to chains. This, in turn,
causes a series of intricate 
bifurcations and a 
loss of stability (see e.g. the second gap in Fig.~\ref{fig6a}). 
In particular, large regions of oscillatory instabilities are born, and hence, 
we expect regions of quasi-periodicity, and of potentially
chaotic response (see e.g. Sec.~\ref{gap2}).
Not surprisingly, the experimentally observed values depart from the 
theoretical 
prediction once stability is lost. However, the general rising and falling pattern
is captured qualitatively, see Fig.~\ref{fig6b}. 
Nevertheless, it is worthwhile 
to note that the latter figure contains a very elaborate
set of loops and a high multiplicity of corresponding (unstable) periodic orbit
families for which no clear physical intuition is apparent at present.
More troubling, however, is 
the fact that there are also some stable regions where the experimental points
are off the theoretical curve. A systematic study using numerical simulations 
revealed that within these narrow stable regions at hand (see e.g.
Fig.~\ref{fig6b}), a large number of periods
is required in order to converge to the exact periodic solution. In particular, 
it happens that the Floquet multipliers are inside the unit circle but very close
to unity, suggesting that the experimental time window of $50\,\textrm{ms}$ used
is not sufficiently long to observe experimentally the periodic structures predicted
theoretically by our model. Thus, the reported experimental observations
 in this regime appear to be capturing a transient stage and not the 
eventual asymptotic form of the dynamical profile. 
A final comment on the possible discrepancy between the 
model and experiments is due to the fact that in this  highly nonlinear 
regime, the
beads come out of contact (the dynamic force exceeds the static force), 
in which 
case other dynamic effects, such as particles' rotations, could have
a significant role, 
which are not described in our simple model~\eqref{gc_start}. In fact, the 
contact points of the particles cannot be perfectly aligned in 
experiments due to the clearance between the beads and the support rods. 
This is likely to 
result in dynamic buckling of the chain under the strong excitations, 
which are considered in some of the cases above. 
Nonetheless, the overall experimental trends of bifurcation in Fig. \ref{fig6a}
are in fair qualitative agreement with the theoretical predictions.

\section{Conclusions \& Future Challenges} \label{theend}
We have presented a natural extension of earlier considerations of (i) 
Hamiltonian breathers in trimer chains and (ii) surface breathers in
damped-driven dimers, by studying a damped-driven trimer granular crystal
through a combination of experimental and theoretical/computational
approaches. We found that the breathers
with frequency in the second gap are in analogy with those in the sole gap
of the damped-driven dimer, in that the interplay of damped-driven dynamics
with nonlinearity and spatial discreteness 
gives rise to saddle-node bifurcations
of time-periodic solutions and turning points beyond which there is
 no stable
ordered dynamics, as well as to surface modes and generally
rather complex and tortuous bifurcation diagrams. 
While similar structures are found in the the first gap, 
they appear to be less robust (given the magnitude of their instabilities)
and can resonate with the higher-order linear bands, a feature interesting 
in its own right. A continuation in driving frequency revealed that the 
nonlinearity causes the resonant peaks to bend, possibility into a spectral 
gap. However, this nonlinear bending also causes the solutions to lose stability
in various ways, such as Neimark-Sacker and saddle-node bifurcations.  
More importantly, all of these features were validated experimentally through
laser Doppler 
vibrometry, allowing, for the first time to our knowledge, the full-field
visualization of surface breathers in granular crystals. 

Several interesting
questions remain unexplored, including  for example the mechanism leading to
the emergence of the observed chaotic dynamics. We also observed that the 
bifurcation structure is generally more complex in the case of breathers that
resonant with the linear modes, which bears further probing. Other possible 
avenues for future research include the investigation of dark breathers in 
such chains. Recall that as in~\cite{dark,dark2}, dark breathers
may bifurcate from the top edge of the acoustic, first or second optical
band, under suitable drive. Lastly, it would be quite interesting
to
generalize considerations of breathers and surface modes in two-dimensional
hexagonal lattices (see e.g.~\cite{leonard} for a recent
example of a study on traveling waves in such settings) 
with both homogeneous, as well as heterogeneous compositions. 
Such studies are currently under consideration and will be
reported in future publications.

\begin{acknowledgments}
E.G.C. gratefully acknowledges financial support from the FP7-People
IRSES-606096: ``\textit{Topological Solitons, from Field Theory to
Cosmos}". P.G.K. acknowledges support from the National Science 
Foundation (NSF) under grants CMMI-1000337, DMS-1312856, from the
ERC and FP7-People 
under grant IRSES-606096 and from the US-AFOSR under grant FA9550-12-1-0332. 
P.G.K.'s work at Los Alamos is supported in part 
by the U.S. Department of Energy. C.C. was partially supported by the ETH Zurich Foundation through the Seed Project ESC-A 06-14.
J.Y. thanks the support of the NSF (CMMI-1414748) and the US-ONR (N000141410388).
E.G.C. and C.C. thank M.O. Williams (PACM, Princeton University) for insight regarding
the AUTO codes used for the bifurcation analysis performed in this work. 
\end{acknowledgments}

\appendix

\section{Numerical methods and linear stability}
\label{sec:app_num_methods}
In this Appendix we shortly discuss the numerical methods employed
for finding exact time-periodic solutions of Eq.~(\ref{gc_start}) together
with the parametric continuation techniques for identifying families
of such solutions as either the value of the actuation frequency
$f_{b}$ (or, equivalently, the period $T_{b}=1/f_{b}$) or amplitude $a$
changes. The interested reader can find more information along 
these directions, e.g., in \cite{ps_method,doedel_I} (among others) and 
references therein.

In order to find time-periodic solutions of Eq.~(\ref{gc_start}), we convert 
the latter into a system of first order ODEs written in compact form
\begin{equation}
\dot{\mathbf{x}}=\mathbf{F}\left(t,\mathbf{u},\mathbf{v}\right),
\label{gc_compact}
\end{equation}
with $\mathbf{x}=\left(\mathbf{u},\mathbf{v}\right)^{T}$, while 
$\mathbf{u}=\left(u_{1},\dots,u_{N}\right)^{T}$ and $\mathbf{v}=\dot{\mathbf{u}}$
represent the $N$-dimensional position and velocity vectors, respectively.
Next, we define the Poincar\'e map: $\mathbf{\cal{P}}(\mathbf{x}^{(0)})=\mathbf{x}^{(0)}-\mathbf{x}(T_{b})$
where $\mathbf{x}^{(0)}$ is the initial condition and $\mathbf{x}(T_{b})$
is the result of integrating Eq.~(\ref{gc_compact}) forward in time until $t=T_{b}$ (using %
e.g., the DOP853 \cite{Hairer} or ODE \cite{Shampine_Gordon} time integrators). Then, a periodic solution with period $T_{b}$ (i.e., satisfying
the property $\mathbf{x}(0)=\mathbf{x}(T_{b})$) will be a root of the map 
$\mathbf{\cal{P}}$. To this end, Newton's method is applied to the map 
$\mathbf{\cal{P}}$ and thus yields the following numerical iteration 
scheme
\begin{equation}
\mathbf{x}^{(0,k+1)}=\mathbf{x}^{(0,k)}-\left[\mathcal{J}\right]^{-1}_{\mathbf{x}^{(0,k)}}\mathbf{\cal{P}}\left(\mathbf{x}^{(0,k)}\right),
\label{gc_newton_iter_scheme}
\end{equation}
together with the Jacobian of the map $\mathcal{P}$, namely, $\mathcal{J}=\mathbf{I}-V(T_{b})$ 
where $\mathbf{I}$ is the $2N\times2N$ identity matrix; $V$ is the solution to the
variational problem: $\dot{V}=\left(D_{\mathbf{x}}\mathbf{F}\right)\,V$ with initial 
data $V(0)=\mathbf{I}$, while $D_{\mathbf{x}}\mathbf{F}$ is the Jacobian of the 
equations of motion~(\ref{gc_compact}) evaluated at the point $\mathbf{x}^{(0,k)}$. 

Subsequently, the iteration scheme (\ref{gc_newton_iter_scheme}) is fed
 a suitable initial guess (for fixed values of the actuation amplitude
$a$ and frequency $f_{b}$) and applied 
until a user-prescribed 
tolerance criterion is satisfied. Thus, a time-periodic solution is 
obtained upon successful convergence (and with high accuracy) together with
the eigenvalues or Floquet multipliers (denoted by $\lambda$) of the {\it monodromy matrix} 
$V(T_{b})$ which convey important information about the stability of the 
solution in question. In particular, a periodic solution is deemed asymptotically
stable (or a stable limit cycle) if there are no Floquet multipliers outside the
unit circle whereas a periodic solution with Floquet multipliers lying outside
the unit circle is deemed to be linearly 
unstable. Note that the iteration scheme (\ref{gc_newton_iter_scheme})
can be initialized using the linear asymptotic equilibrium given by Eq.~(\ref{steady_state_lin_ansatz}).
Alternatively, one may start with zero initial data and directly integrate
the equations of motion (\ref{gc_compact}) forward in time. Then, 
Eq.~(\ref{gc_newton_iter_scheme}) can be initialized with the output waveform of the 
time integrator at a particular time. 

Having obtained an exact periodic solution to Eq.~(\ref{gc_start}) for given
values of $a$ and $f_{b}$, we perform parametric continuations over these 
parameters using the computer software AUTO \cite{AUTO}. This way, we are 
able to trace entire families of periodic solutions and study their corresponding
stability characteristics in terms of the Floquet multipliers.


\begin{thebibliography}{99}


\bibitem{Nesterneko_book} V.F. Nesterenko, \textit{Dynamics of Heterogeneous Materials}, %
                  (Springer-Verlag, New York, 2001).   

\bibitem{sen08} S. Sen, J. Hong, J. Bang, E. Avalos, and R. Doney, Phys. Rep. {\bf 462}, 21 (2008).

\bibitem{Theocharis_rev} G. Theocharis, N. Boechler, and C. Daraio,
in {\it Phononic Crystals and Metamaterials}, Ch. 6, 
Springer Verlag,
(New York, 2013)

\bibitem{coste} C. Coste, E. Falcon, and S. Fauve, Phys. Rev. E {\bf 56}, 6104 (1997).

\bibitem{pik} K. Ahnert and A. Pikovsky, Phys. Rev. E {\bf 79}, 026209 (2009).

\bibitem{herbold} E. B. Herbold and V. F. Nesterenko,
Phys. Rev. E {\bf 75}, 021304 (2007).

\bibitem{molin} A. Molinari and C. Daraio, Phys. Rev. E {\bf 80}, 056602 (2009).
           
\bibitem{Theo2009} G. Theocharis, M. Kavousanakis, P.~G. Kevrekidis, C. Daraio, M.~A. Porter, and I.~G. Kevrekidis, Phys. Rev. E {\bf 80}, 066601 (2009);
S. Job, F. Santibanez, F. Tapia and F. Melo, Phys. Rev. E {\bf 80}, 
025602 (2009);
Y. Man, N. Boechler, G. Theocharis, P. G. Kevrekidis, and C. Daraio,
Phys. Rev. E {\bf 85}, 037601 (2012).

\bibitem{Theo2010} N. Boechler, G. Theocharis, S. Job, P.~G. Kevrekidis, M.~A. Porter and C. Daraio, Phys. Rev. Lett. {\bf 104}, 244302 (2010);
G. Theocharis, N. Boechler, P.~G. Kevrekidis,  S. Job, M.~A. Porter, and C. Daraio, Phys. Rev. E {\bf 82}, 056604 (2010). 

\bibitem{Nature11} N. Boechler, G. Theocharis, C. Daraio,  Nature Materials \textbf{10}, 665 (2011). 

\bibitem{hooge12} C. Hoogeboom, Y. Man, N. Boechler, G. Theocharis, P. G. Kevrekidis, I. G. Kevrekidis and C. Daraio, Euro. Phys. Lett. {\bf 101 }, 44003 (2013).   

  \bibitem{hooge13}  C. Hoogeboom and P. G. Kevrekidis,
Phys. Rev. E {\bf 86}, 061305 (2012).


\bibitem{dark} C. Chong, P.G. Kevrekidis, G. Theocharis, and C. Daraio, 
Phys. Rev. E. \textbf{87}  042202 (2013).      

\bibitem{dark2} C. Chong, F. Li, J. Yang, M.O. Williams, I.G. Kevrekidis,   %
           P.G. Kevrekidis and C. Daraio, Phys. Rev. E, \textbf{89}, 032924 (2014). 




\bibitem{dar06} C. Daraio, V.~F. Nesterenko, E.~B. Herbold, and S. Jin, Phys. Rev. Lett. {\bf 96}, 058002 (2006).

\bibitem{hong05} J. Hong, Phys. Rev. Lett. {\bf 94}, 108001 (2005).

\bibitem{fernando} F. Fraternali, M.~A. Porter, and C. Daraio, Mech. Adv. Mat. Struct. {\bf 17}(1), 1 (2010).


\bibitem{doney06} R. Doney and S. Sen, Phys. Rev. Lett. {\bf 97}, 155502 (2006).

\bibitem{dev08} D. Khatri, C. Daraio, and P. Rizzo, SPIE {\bf 6934}, 69340U (2008).



\bibitem{Spadoni} A. Spadoni and C. Daraio, Proc Natl Acad Sci USA, {\bf 107}, 7230, (2010).

\bibitem{Li_switch} F. Li, P. Anzel, J. Yang, P.G. Kevrekidis, and C. Daraio, Nature Communications, in press (2014).




\bibitem{dar05} C. Daraio, V.~F. Nesterenko, E.~B. Herbold, and S. Jin, Phys. Rev. E {\bf 72}, 016603 (2005).

\bibitem{dar05b} V.~F. Nesterenko, C. Daraio, E.~B. Herbold, and S. Jin, Phys. Rev. Lett. {\bf 95}, 158702 (2005).


\bibitem{Flach2007} S. Flach and A.~V. Gorbach, Phys. Rep. {\bf 467}, 1 (2008).




\bibitem{moti} 
F. Lederer, 
G.I. Stegeman, D.N. Christodoulides, G. Assanto,
M. Segev, and Y. Silberberg,
Phys. Rep. {\bf 463}, 1 (2008).

\bibitem{sievers}
M. Sato, 
B.E. Hubbard, and A.J. Sievers,
Rev. Mod. Phys.  {\bf 78}, 137 (2006).

\bibitem{alex} 
P. Binder, 
D. Abraimov, A.V. Ustinov, S. Flach, and Y. Zolotaryuk,
Phys. Rev. Lett. {\bf 84}, 745 (2000); 
\\
E. Tr{\'{\i}}as, 
J.J.  Mazo, and T.P. Orlando,
Phys. Rev. Lett. {\bf 84}, 741 (2000).


\bibitem{lars3}
L.Q. English, 
M. Sato, and A.J. Sievers,
Phys. Rev. B {\bf 67}, 024403 (2003); 
\\
U.T. Schwarz, 
L.Q. English, and A.J. Sievers,
Phys. Rev. Lett. {\bf 83}, 223 (1999).

\bibitem{swanson}
B.I. Swanson, 
J.A. Brozik, S.P. Love, G.F. Strouse, A.P. Shreve,
A.R. Bishop, W.-Z. Wang, and M.I. Salkola, 
Phys. Rev. Lett. {\bf 82}, 3288 (1999).

\bibitem{Peybi} 
M. Peyrard, Nonlinearity {\bf 17}, R1 (2004).

\bibitem{Morsch} O. Morsch and M. Oberthaler, Rev. Mod. Phys. {\bf 78}, 179
(2006).



\bibitem{James01} G. James, C.R. Acad. Sci.,Ser.1:Math {\bf 332}, 581 (2001).

\bibitem{James03} G. James, J. Nonlinear Sci. \textbf{13}, 27 (2003).


\bibitem{man}   Y. Man, N. Boechler, G. Theocharis, P. G. Kevrekidis, and C. Daraio, Phys. Rev. E {\bf 85}, 037601 (2012).
     
        
 \bibitem{mason1}  M. A. Porter, C. Daraio, E. B. Herbold, I. Szelengowicz, and P. G. Kevrekidis, Phys. Rev. E {\bf 77}, 015601 (2008).
\bibitem{mason2} M. A. Porter, C. Daraio, I. Szelengowicz, E. B. Herbold, and P. G. Kevrekidis, Physica D {\bf 238}, 666 (2009).
     
 
\bibitem{Tun_Band_gaps} N. Boechler, J. Yang, G. Theocharis, P.G. Kevrekidis and   %
                        C. Daraio, J. Appl. Phys., \textbf{109}, 074906 (2011).
     

\bibitem{Hertz} H. Hertz,
J. Reine Angew. Math., \textbf{92} (1881), 156--171.
           
\bibitem{Johnson_contact} K.L. Johnson, \textit{Contact Mechanics} (Cambridge
University Press, Cambridge, 1985).


\bibitem{tristram} Y. Xu, T.J. Alexander, H. Sidhu and P.G. Kevrekidis,
Phys. Rev. E {\bf 90}, 042921 (2014).

\bibitem{leonard} A. Leonard, C. Chong, P. G. Kevrekidis, and C. Daraio,
Granular Matter {\bf 16}, 531 (2014).



\bibitem{ps_method} A.H. Nayfeh and B. Balachandran, \textit{Applied
Nonlinear Dynamics: Analytical, Computational and Experimental Methods}
(Willey Series in Nonlinear Science, 1995).                   
                   
\bibitem{doedel_I} E. Doedel, H.B. Keller and J.P. Kern\'evez, Internat. %
                   J. Bifur. and Chaos, \textbf{01}, 493-520 (1991).

\bibitem{Hairer} E. Hairer, S.P. N{\o }rsett and G. Wanner, \textit{Solving
                 ordinary differential equations I} (Springer-Verlag, Berlin, 1993).

\bibitem{Shampine_Gordon} L. Shampine and M. Gordon, \textit{Computer Solution %
                 of Ordinary Differential Equations: The Initial Value Problem}, %
                 (W.H.Freeman \& Co Ltd, 1975).
                   
\bibitem{AUTO} E. Doedel, AUTO, \textit{indy.cs.concordia.ca/auto/}. 

\end{thebibliography}
\end{document}